\documentclass[sigconf]{acmart}

\usepackage{amsmath,amsfonts}

\usepackage{amsthm}

\newtheorem*{prob_state}{\textbf{Problem Statement}}

\newtheorem{theorem}{\textbf{Theorem}}

\usepackage{float}

\usepackage{color}
\usepackage{tikz}
\usetikzlibrary{trees}
\usepackage{amsthm}
\usepackage{makecell}
\usepackage{tikz}
\usetikzlibrary{trees}
\usepackage{tabularx,colortbl}
\usepackage{threeparttable}
\usepackage{booktabs}
\usepackage{multirow}
\usepackage{subfig}
\usepackage{graphicx}
\usepackage{url}
\usepackage{enumitem}
\usepackage{tcolorbox}
\usepackage{xcolor}
\usepackage[ruled, vlined, linesnumbered]{algorithm2e}
\usepackage{hyperref}
\usepackage[noabbrev]{cleveref}
\crefname{equation}{Eq.}{Eqs.}

\newcommand{\filledcircle}{\tikz\fill[black] (0,0) circle (.8ex);}
\newcommand{\emptycircle}{\tikz\draw (0,0) circle (.8ex);}

\definecolor{DeepPink}{HTML}{FF1493}
\definecolor{Orchid}{HTML}{DA70D6}
\definecolor{Magenta}{HTML}{FF00FF}
\definecolor{Fuchsia}{HTML}{FF00FF}
\definecolor{LavenderPink}{HTML}{FFB6C1}
\definecolor{verylightgray}{rgb}{0.9, 0.9, 0.9}
\definecolor{lightred}{rgb}{1,0.8,0.8}

\AtBeginDocument{%
  }

\copyrightyear{2025}
\acmYear{2025}
\setcopyright{rightsretained}
\acmConference[WWW '25]{Proceedings of the ACM Web Conference 2025}{April 28-May 2, 2025}{Sydney, NSW, Australia}
\acmBooktitle{Proceedings of the ACM Web Conference 2025 (WWW '25), April 28-May 2, 2025, Sydney, NSW, Australia}

\makeatletter
\gdef\@copyrightpermission{
	\begin{minipage}{0.2\columnwidth}
		\href{https://creativecommons.org/licenses/by/4.0/}{\includegraphics[width=0.90\textwidth]{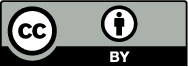}}
	\end{minipage}\hfill
	\begin{minipage}{0.8\columnwidth}
		\href{https://creativecommons.org/licenses/by/4.0/}{This work is licensed under a Creative Commons Attribution International 4.0 License.}
	\end{minipage}
	\vspace{5pt}
}
\makeatother

\begin{document}

\title{TAPE: Tailored Posterior Difference for \\Auditing of Machine Unlearning}
\subtitle{ \it \textbf{ To appear at The Web Conference 2025. Author version} }


\author{Weiqi Wang}
\email{Weiqi.Wang@uts.edu.a}
\orcid{0000-0002-7905-3126}
\affiliation{%
	\institution{University of Technology Sydney}
	\city{Sydney}
	\state{NSW}
	\country{Australia}}

\author{Zhiyi Tian}
\email{Zhiyi.Tian-1@uts.edu.au}
\authornote{Corresponding author: Zhiyi Tian. \\ \textbf{This paper is the draft that preparing for publishing on WWW25.}}
\orcid{0000-0001-8905-0941}
\affiliation{%
	\institution{University of Technology Sydney}
	\city{Sydney}
	\state{NSW}
	\country{Australia}
}

\author{An Liu}
\email{anliu@suda.edu.cn}
\orcid{0000-0002-6368-576X}
\affiliation{%
	\institution{Soochow University}
	\city{Soochow}
	\state{Jiangsu}
	\country{China}
}
\author{Shui Yu}
\email{shui.yu@uts.edu.au}
\orcid{0000-0003-4485-6743}
\affiliation{%
	\institution{University of Technology Sydney}
	\city{Sydney}
	\state{NSW}
	\country{Australia}
}

\renewcommand{\shortauthors}{WeiqiWang, Zhiyi Tian, An Liu, and Shui Yu}


\begin{abstract}
With the increasing prevalence of Web-based platforms handling vast amounts of user data, machine unlearning has emerged as a crucial mechanism to uphold users' right to be forgotten, enabling individuals to request the removal of their specified data from trained models. 
However, the auditing of machine unlearning processes remains significantly underexplored. Although some existing methods offer unlearning auditing by leveraging backdoors, these backdoor-based approaches are inefficient and impractical, as they necessitate involvement in the initial model training process to embed the backdoors. In this paper, we propose a TAilored Posterior diffErence (TAPE) method to provide unlearning auditing independently of original model training. We observe that the process of machine unlearning inherently introduces changes in the model, which contains information related to the erased data. TAPE leverages unlearning model differences to assess how much information has been removed through the unlearning operation. Firstly, TAPE mimics the unlearned posterior differences by quickly building unlearned shadow models based on first-order influence estimation. Secondly, we train a Reconstructor model to extract and evaluate the private information of the unlearned posterior differences to audit unlearning. Existing privacy reconstructing methods based on posterior differences are only feasible for model updates of a single sample. To enable the reconstruction effective for multi-sample unlearning requests, we propose two strategies, unlearned data perturbation and unlearned influence-based division, to augment the posterior difference. Extensive experimental results indicate the significant superiority of TAPE over the state-of-the-art unlearning verification methods, at least 4.5$\times$ efficiency speedup and supporting the auditing for broader unlearning scenarios. 



\end{abstract}


\maketitle

\section{Introduction} \label{intro}

Rising concerns over personal data privacy have led to the enactment of stringent privacy regulations and laws, such as the General Data Protection Regulation (GDPR)~\cite{mantelero2013eu}. These legal frameworks guarantee individuals the ``right to be forgotten'', granting the right to request the removal of their data when participating in machine learning (ML) services. This right has sparked significant interest in the research community, giving rise to the concept of ``machine unlearning'' --- a field that explores methods for erasing the influence of user-specified samples from trained ML models~\cite{bourtoule2021machine,neel2021descent,warnecke2024machine}. For example, in Web-based recommendation systems that collect huge amounts of sensitive user data, effective unlearning methods are essential for protecting user privacy \citep{liu2024breaking,chen2022recommendation}. Although many unlearning techniques are proposed, most of them focus on developing unlearning optimization algorithms while ignoring the provision of unlearning auditing.


\noindent
\textbf{Research Gap.}
There are a few works provided unlearning execution verification based on backdoor techniques~\cite{hu2022membership,guo2023verifying,sommer2022athena}. However, the backdoor-based methods have two oblivious disadvantages: (1) they are inefficient in practice as they are required to backdoor the model in the original model training period; (2) they cannot provide exact verification for genuine samples.

\begin{figure}[t]
	\centering
	\includegraphics[width=1.0\linewidth]{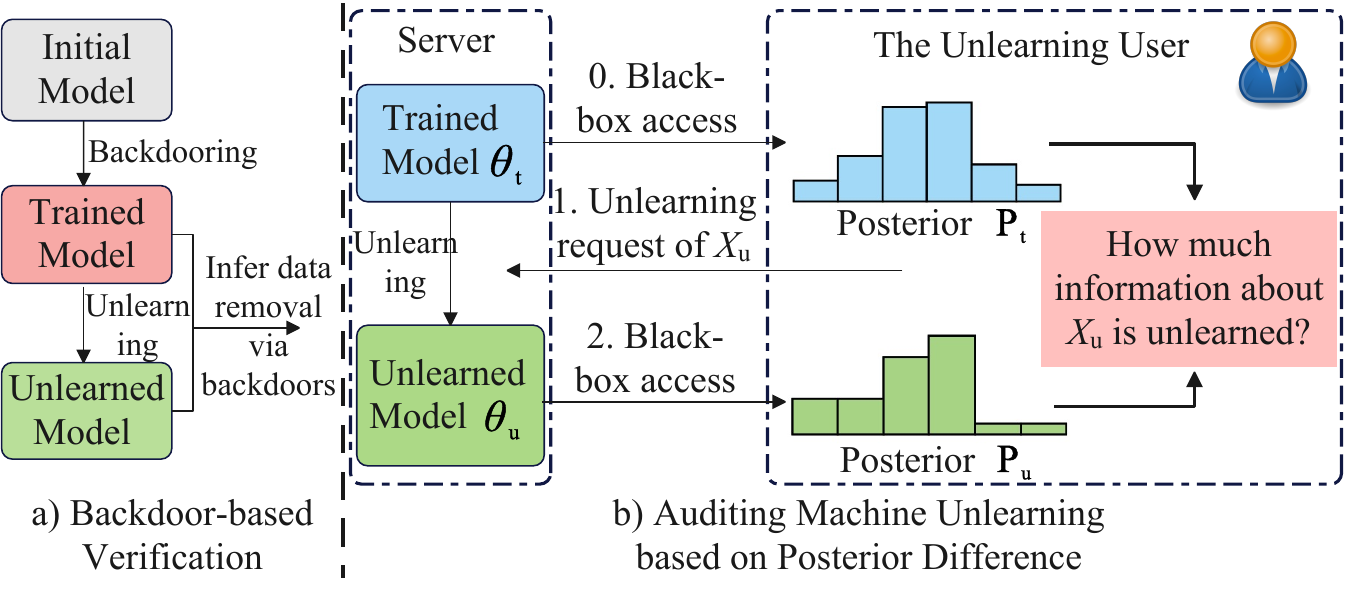} 
	\caption{(a) The backdoor-based verification and (b) The motivation of auditing unlearning effectiveness based on the posterior difference. The scheme (b) only involves the unlearning process rather than the initial model training. \vspace{-4mm}
	}
	\label{fig_unlearningauditfigure1}
\end{figure}

First, the efficacy of backdoor-based unlearning verification schemes hinges on model backdooring during the initial model training process \cite{hu2022membership,guo2023verifying}, as shown in~\Cref{fig_unlearningauditfigure1}(a), which is impractical and inefficient. Users are unlikely to foresee the need to unlearn specific samples at the outset, making it unreasonable to incorporate tailored backdoors for specified samples during the initial training phase. Furthermore, involving the model training process in this way introduces inefficiencies, as it would be more effective to design an audit method that focuses solely on the machine unlearning operation. 

Second, the backdooring method can only build the connection between backdoored samples and models, but the backdoored samples and erased genuine samples are distinct datasets~\cite{gao2023backdoor,pan2023asset}. These two datasets behave differently during model training, especially in approximate unlearning \cite{nguyen2020variational,fu2022knowledge}, where the model accuracy on backdoored samples diminishes much faster than that on genuine samples~\cite{wang2019neural,pan2023asset}. 
It indicates that the removal of backdoors can only verify whether the backdoored samples are unlearned from the model rather than genuine samples.

\noindent
\textbf{Research Question.}
Based on the research gap, we pose the research question: \textit{``When an unlearning request is uploaded and processed, can we provide a practical audit service that verifies data removal and assesses the effectiveness of unlearning?''} Specifically, for practicality, the audit should only involve the unlearning process, and for effectiveness, it should rigorously determine whether the specified data has been unlearned and evaluate how much information has been erased from the model. 

\noindent
\textbf{Motivation.}
The process of machine unlearning inevitably results in two versions of the model: one before and one after unlearning \cite{chen2021machine}. The difference between these models encapsulates the privacy information of the erased samples \cite{Hu2024sp,chen2021machine,zhang2023conditional}. Our approach to audit unlearning effectiveness is based solely on analyzing these model differences, as illustrated in~\Cref{fig_unlearningauditfigure1}(b).
Auditing unlearning effectiveness based on model differences offers two key advantages. First, this method is practical and efficient, as it focuses exclusively on the unlearning operations without requiring involvement in the initial model training process. Second, auditing based on model differences supports broader unlearning scenarios and requests because unlearning for either genuine or backdoored samples results in model differences. By contrast, backdoor-based methods are only effective for backdoored samples as they can only build connections between backdoored samples and models.

\noindent
\textbf{Our Work.}
In this paper, we address the research question by formalizing the machine unlearning auditing problem and introducing an approach called TAPE, designed to audit unlearning effectiveness solely based on the unlearning process. TAPE contributes one method and two strategies to effectively train a Reconstructor model to evaluate how much private information is unlearned to audit this unlearning update. As a preparatory step, TAPE mimics unlearning posterior differences as input data for the Reconstructor by proposing an unlearned shadow model establishment method based on first-order influence estimation. While existing privacy reconstruction methods are effective for single samples, they are impractical for multiple samples--a common scenario in unlearning requests. To address this, we leverage the fact that the unlearning user knows and uploads the erased data, allowing us to design two strategies to ensure our method is suitable for multi-sample scenarios. Specifically, we design the unlearned data perturbation strategy to augment the posterior difference for a better reconstruction effect of unlearned samples. Additionally, we develop an unlearned influence-based division strategy, which transforms the reconstruction task from dealing with multiple samples as a single posterior difference to reconstructing each sample individually based on multiple divided posterior differences, significantly enhancing the overall reconstruction effectiveness.

We conduct extensive experiments on four representative datasets and four mainstream unlearning benchmarks to evaluate the proposed method, in which the results indicate the superiority of TAPE over the start-of-the-art auditing methods \cite{hu2022membership,guo2023verifying} in terms of both efficiency and efficacy. From the efficiency perspective, our TAPE method achieves at least $4.5\times$ speedup on all datasets and at most $75\times$ speedup on the CelebA dataset than backdoor-based methods, as TAPE only involves the unlearned process. In contrast, backdoor-based methods must backdoor the service model during the initial training process, which is computationally expensive. From the efficacy perspective, our TAPE provides effective auditing of genuine samples for both exact and approximate unlearning algorithms, while the verification of backdoor-based methods only targets backdoored samples.

Our contributions are summarized as follows:
\begin{itemize}[itemsep=0pt, parsep=0pt, leftmargin=*]
	\item This paper is the \textit{first} to investigate the auditing for machine unlearning involves only the unlearning process, which is much different from existing backdoor-based methods that rely on backdooring the model during the initial training process. Moreover, our auditing study is feasible for genuine unlearned samples rather than only backdoor-marked samples. 
	\item We propose a TAPE method based on the posterior difference to auditing unlearning. TAPE introduces a novel method to quickly establish unlearned shadow models that mimic the posterior differences and incorporates two posterior augmentation strategies to facilitate auditing the unlearning of multiple samples. 
	\item We conduct extensive experiments on both exact and approximate unlearning methods across representative datasets and various model architectures. The findings validate significant improvements in efficiency and broader applicability to adaptive unlearning scenarios compared with the state-of-the-art unlearning verification methods. 
	\item The source code and the artifact of TAPE is released at \url{https://anonymous.4open.science/r/TAPE-30D0}, which creates a new tool for measuring the effectiveness of machine unlearning methods, shedding light on the design of future unlearning auditing methods. 
\end{itemize}


\section{Related Work} \label{related_work}

Few backdoor-based solutions \cite{hu2022membership,sommer2022athena,guo2023verifying,gao2024verifi} provided data removal verification for machine unlearning. These studies mixed backdoored samples to users' data for backdooring servers' service ML models during the model training process. Then, they inferred whether the users' data was unlearned by testing if the backdoor disappeared from the service models \cite{hu2022membership,guo2023verifying}.
However, these backdoor-based methods have two oblivious limitations. 

\begin{figure}[t]
	\centering
	\includegraphics[width=0.87\linewidth]{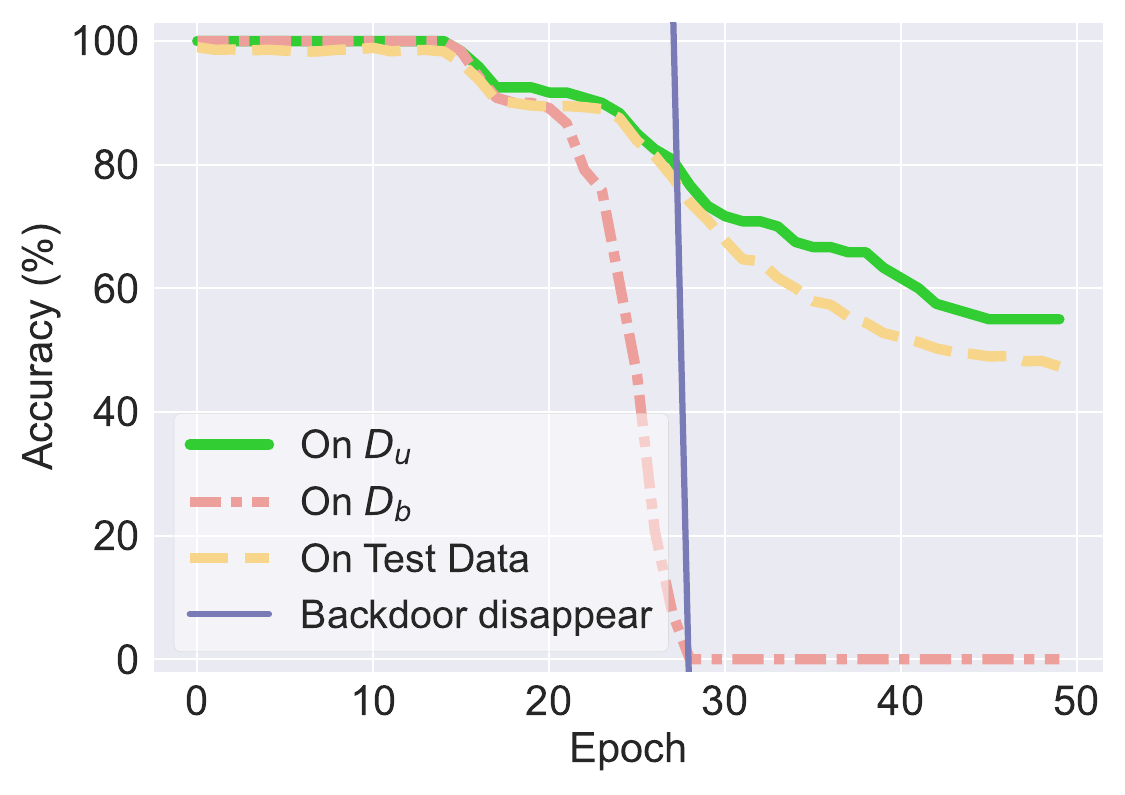}
	\vspace{-2mm}
	\caption{Approximate unlearning process on genuine unlearned data $D_u$ and backdoored data $D_b$ on MNIST. During unlearning, the backdoor accuracy drops to 0\% at the blue Vertical line. Meanwhile, the model accuracy on genuine unlearned data $D_u$ and test data is still around $80\%$.
	}
	\vspace{-2mm} 
	\label{fig_mnistepochaccdrop}
\end{figure}

First, these methods rely on the original ML model training process, which is impractical in real-world scenarios due to users being unaware of which samples will need to be unlearned in the future, as well as the high computational costs involved. Second, these methods only build the connection between the backdoored samples and models, while the backdoored dataset and genuine dataset are still separate from the models' perspective~ \cite{wang2019neural,zeng2023narcissus}. The backdoored and the erased genuine datasets perform differently during model training, as the corresponding experimental results are shown in \Cref{fig_mnistepochaccdrop}. The removal of backdoor triggers can actually only verify whether the backdoored samples are unlearned as the backdoored samples and unlearned genuine samples perform differently during the approximate unlearning training process \cite{nguyen2020variational}. When the accuracy of backdoored samples drops to $0\%$, the model accuracy on genuine unlearned data and test data is still around $80\%$. These results are consistent with current backdoor studies \cite{gao2023backdoor,pan2023asset,wang2023mm}. We present more detailed discussion in \Cref{different_with_existing}.

\section{Preliminary and Problem Statement} \label{problem_df}



%


To facilitate the understanding of the unlearning auditing problem, we first introduce the main process of unlearning. A detailed introduction about the 
 and threat model is presented in \Cref{threat_model}. 

\noindent
\textbf{Machine Unlearning.} The unlearning process usually includes the following phases. (1) The server trained a model with parameters $\theta_t$ derived from dataset $D$. (2) The unlearning user uploads the unlearning requested dataset $D_u$ to the server for unlearning. (3) The server conducts an unlearning algorithm $\mathcal{U}$ to remove $D_u$'s contribution from $\theta_t$ and results in an unlearned model with parameters $\theta_{u, D \backslash D_u}$, also denoted as $\theta_u$. 

Most existing backdoor-based unlearning verification methods tried to solve data removal verification but can only answer if the backdoored samples are unlearned. Answering whether the backdoored data is or not deleted is insufficient for trustworthy unlearning auditing. We should assess the unlearning effectiveness of the model, i.e., how much private information about the requested unlearning samples is removed from the model.

\begin{prob_state}[Unlearning Effectiveness Audit] 
	\label{effectiveness_problem}	
Given the described unlearning scenario, the potential for unlearning execution spoofing by the server, and the capabilities of the unlearning user, auditing unlearning effectiveness necessitates a method for unlearning users to evaluate the extent to which information about $D_u$ has been unlearned from $\theta_t$ to $\theta_{u}$.
\end{prob_state}

%

It is important to note that the problem statement inherently includes the issue of data removal verification. If one can effectively measure how much information related to the erased samples has been unlearned, this measurement can serve as the basis for determining whether the data has been properly unlearned. 
We try to conduct unlearning auditing based on the unlearning updated posterior difference as it contains essential information about the erased samples. To achieve the auditing goals, we need to mimic the unlearning posterior difference and extract and quantify the unlearned information from it. We utilize the model's output layer results of the original and unlearned models on the user's local dataset to generate the posterior difference. 
We define the unlearning posterior difference as follows.

\noindent
\textbf{Posterior Difference.}
The unlearning user first queries the trained ML model $\theta_t$ before unlearning with all samples of $D_{local}$ and concatenates the received outputs to form a vector $\hat{Y}_{t, local}$. Then, the user queries the unlearned model $\theta_u$ with samples in the $D_{local}$ and creates a vector $\hat{Y}_{u, local}$. In the end, the user sets the posterior difference, denoted by $\delta$, to the difference of both outputs: 
\begin{equation} \label{posterior_diff}
	\delta = \hat{Y}_{t,local} - \hat{Y}_{u,local}.
\end{equation}
Note that the dimension of $\delta$ is the product of $D_{local}$'s cardinality and the number of classes of the target dataset. For example, in this paper, CIFAR-10 and MNIST are 10-class datasets, while we just identify the gender attributes of CelebA, which is a binary classification. As we set the local dataset $0.5\%$ of CIFAR-10 and MNIST, and $0.06\%$ of CelebA, this indicates the dimension of $\delta$ is 2500 for CIFAR-10, 3000 for MNIST, and 1210 for CelebA.

\begin{figure*}[t]
	\centering
	\includegraphics[width=0.97\linewidth]{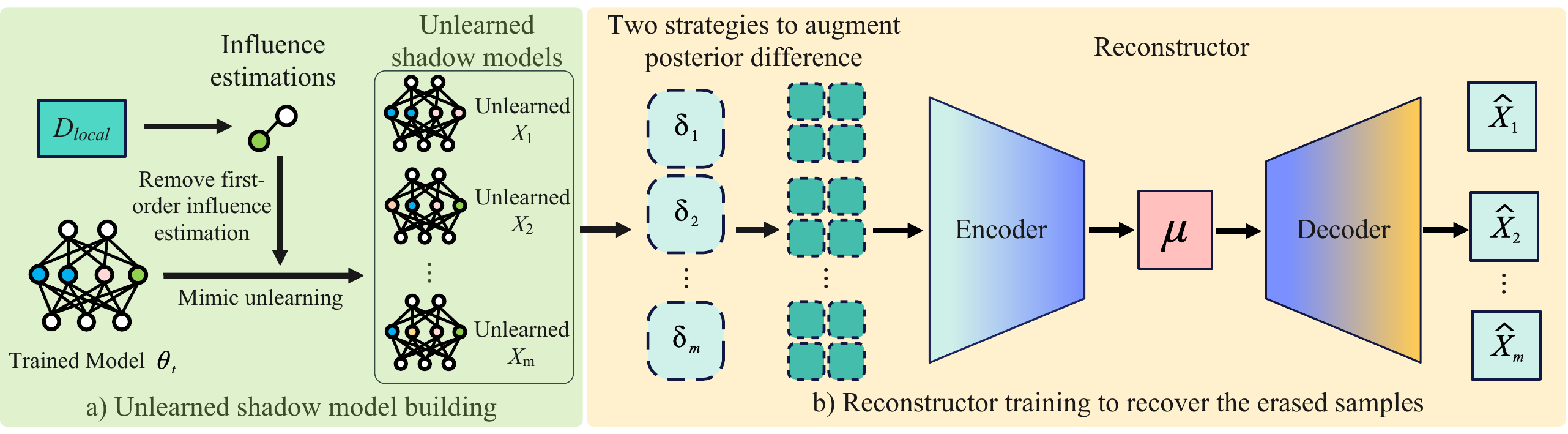}
	\vspace{-2mm}
	\caption{The main process of the TAPE method. (a) The first part quickly builds the unlearned shadow models through first-order influence estimation based on the user's local dataset $D_{local}$ to mimic the unlearning posterior difference $\delta$. (b) Two posterior difference augment strategies are proposed to make the reconstruction suitable for multi-sample unlearning. \vspace{-2mm}}
	\label{fig_reconstructionattack}
\end{figure*}

\noindent
\textbf{Unlearned Information Reconstruction to Assess How Much Information is Unlearned.}
To assess the unlearning effectiveness, we employ a reconstructor model to extract the unlearned information from the posterior difference. We employ the cosine similarity between the reconstructed and original unlearned samples to assess how much information of the unlearned information can be recovered from the unlearning update:  
\begin{equation} \label{similarity_eq}
	\textbf{Rec. Similarity:} \hspace{12mm} \text{sim}(\hat{X}_{u}, X_u) = \frac{\hat{X}_{u} \cdot X_u}{ \| \hat{X}_{u} \| \cdot \| X_u \|}.
\end{equation}
Here, $\hat{X}_{u} \cdot X_u$ is the dot product of the reconstructed vectors $\hat{X}_{u}$ and original unlearned samples vectors $X_u$. $\| \hat{X}_{u}\|$ and $\|X_u\|$ are the Euclidean norms of the two vectors. A higher reconstruction similarity means more information about the erased samples is unlearned from the model.

\section{TAPE Methodology} \label{muv_method}


\subsection{Overview of the TAPE}


We illustrate the overview methodology process in \Cref{fig_reconstructionattack}, which includes two main steps. 


\noindent
\textbf{Unlearned Shadow Model Building.} In this step, we propose a method to quickly build the unlearned shadow models with only the user's samples. Our method utilizes the first-order influence estimation function to effectively estimate the unlearning influence and remove it from the original model, thus quickly mimicking the unlearned model to generate the posterior differences.

\noindent
\textbf{Reconstructor Training.} We then train a reconstructor model to evaluate how much information about the erased samples is unlearned. An unlearned data perturbation strategy and an unlearned influence-based division strategy are proposed to augment the posterior differences for reconstruction for multiple samples. Both strategies are implemented utilizing the advantage that the unlearning user knows and prepares the unlearned samples.




\subsection{Constructing Unlearned Shadow Model to Mimic Posterior Difference} \label{fast_g}

The unlearning user possesses a local dataset $D_{local}$, including the unlearned data $D_u$. 
As the unlearning verification is executed on the unlearning user side, the user can utilize the local dataset $D_{local}$ to construct unlearned shadow models to mimic posterior differences.  

\noindent
\textbf{Constructing Unlearned Shadow Model.} Many existing machine unlearning algorithms rely on the assistance of the remaining dataset $D \backslash D_u$. We propose a method based on the influence function theory \cite{koh2017understanding,basu2020second,bae2022if} in ML to quickly approximate an unlearned shadow model with only the unlearned data $D_u$. Specifically, when we remove $D_u$ from a trained model $\theta_t$ for unlearning, the empirical risk minimization (ERM) can be written as:
\begin{equation}
	\small
	\mathcal{L}_{D \backslash D_u}(\theta) = \frac{1}{n - m} \sum_{x \in D \backslash D_u} \ell (x; \theta),
\end{equation}
where $n$ is the size of $D$, $m$ is the size of $D_u$, and $\ell(x;\theta)$ is the loss. 

Similar to \cite{basu2020second}, we evaluate the effect of up-weighting a group of training samples on model parameters. Note that in this case, the updated weights must still form a valid distribution. Specifically, if a group of training samples is up-weighted, the weights of the remaining samples should be down-weighted to preserve the sum to one constraint of weights in the ERM formulation. 
We assume that the weights of samples in $D_u$ have been up-weighted all by $\epsilon$ and use $\frac{m}{n}$ to denote the fraction of up-weighted training samples. This results in a down-weighting of the rest of the training data by $\tilde{\epsilon} = \frac{m}{n-m} \epsilon$, to preserve the empirical weight distribution of the training dataset. Then, the ERM can be translated as:
\begin{equation} \label{loss_of_unlearning}
	\mathcal{L}^{\epsilon}_{D \backslash D_u} (\theta) = \frac{1}{n} (\sum_{x \in D \backslash D_u} (1 - \tilde{\epsilon})\ell(x;\theta) + \sum_{x \in D_u}(1+\epsilon)\ell(x;\theta)).
\end{equation}
In the above equation, if $\epsilon=0$, we get the original loss function $\mathcal{L}_{\emptyset}(\theta)$ (none of the training data points are unlearned) and if $\epsilon=-1$, we get the loss function $\mathcal{L}_{D \backslash D_u}(\theta)$ (specified samples are removed).


Let $\theta^{\epsilon}_{D \backslash D_u}$ denote the optimal parameters for $\mathcal{L}^{\epsilon}_{D \backslash D_u}$ minimization, and $\theta^*$ denote the optimal parameters trained on $D$. The unlearned shadow models can be approximately achieved by removing the estimated data influence from the trained model as follows,
\begin{equation} \label{shadow_model}
	\theta^{\epsilon}_{D \backslash D_u}  =  \theta_t  -  \frac{\epsilon}{n-m} \sum_{x_u \in D_u} \nabla \ell (x_u; \theta_t),
\end{equation}
where $\epsilon \in [-1,0]$ is used for unlearning, $m$ is the size of the erased dataset and $n$ is the size of the training dataset. $\Delta \theta \simeq  - \frac{\epsilon}{n-m} \sum_{x_u \in D_u} \nabla \ell (x_u; \theta_t)$ is the estimaed data influence at current trained model $\theta_t$. We omit the proof of the shadow model estimation in \Cref{shadow_model} as it is similar to the proofs in \cite{koh2017understanding,basu2020second}. Constructing the unlearned shadow model based on \Cref{shadow_model} only relies on the unlearned samples and is convenient for the user to implement. 

\noindent
\textbf{Mimicking Posterior Difference.} 
With the above method to construct unlearned shadow models, then, we can easily achieve the mimicked posterior differences. For instance, assuming the local dataset $D_{local}$ contains $m$ samples $X_1, X_2, ..., X_m$, we can construct $m$ unlearned shadow models $\theta_{D \backslash X_1}, \theta_{D \backslash X_2}, ..., \theta_{D \backslash X_m}$ for each sample, where $\backslash X$ means unlearning the sample $X$. Based on these unlearned shadow models, we can mimic the corresponding unlearned posterior, $\hat{Y}_{\backslash X_1, local}, \hat{Y}_{\backslash X_2, local}, ..., \hat{Y}_{\backslash X_m, local}$, using the local dataset. The posterior difference can be calculated through \Cref{posterior_diff}, denoted as $\delta_1, \delta_2, ..., \delta_m$, as shown in \Cref{fig_reconstructionattack}. Together with the corresponding shadow unlearning set's ground truth information, the training data for the reconstructor model to evaluate the unlearned information is derived.


\subsection{Reconstructor Model Training with Two Strategies for Multiple Samples Auditing} \label{two_s}
 

\noindent
\textbf{Reconstructor Training for Unlearning Effectiveness Assessment.} Like \citep{salem2020updates}, we employ the autoencoder ($\texttt{AE}$) architecture to construct the Reconstructor, which includes an encoder and a decoder, as shown in \Cref{fig_reconstructionattack}(b). Its goal is to learn an efficient encoding for the posterior differences $\delta$. The encoder encodes the posterior difference into a latent vector $\mu$, and the decoder decodes the latent vector to reconstruct the unlearned samples. We employ mean squared error (MSE) as the loss function, $\mathcal{L}_{\texttt{AE}} = ||\hat{X_u} - X_u||^2_2,$ where $\hat{X_u} = \texttt{AE} (\delta_u)$ is the reconstructed sample for $X_u$. 

Existing studies \cite{salem2020updates, Hu2024sp, balle2022reconstructing} showed effective reconstruction for a single sample for the updated model difference. However, they are infeasible for reconstructing multiple samples. For unlearning effectiveness auditing, the unlearning user has the knowledge of the unlearned samples. With this advantage, we design two strategies: one augments posterior differences by perturbing unlearned data before unlearning and one augments posterior differences by individually dividing the posterior difference after unlearning, enabling evaluate how much information is unlearned for multiple samples.



\noindent
\textbf{Unlearned Data Perturbation before Unlearning.} 
We propose an unlearned data perturbation method to augment posterior difference, assisting the unlearning effectiveness verification. Specifically, we hope to introduce a perturbation $\Delta^p$ to the unlearned sample $X_u$ to augment the unlearned posterior for the reconstructor, so that it can effectively evaluate how much information is unlearned. At the same time, unlearning the perturbed specified data should maintain the unlearned model's utility on the remaining dataset. Since our final purpose is to improve the reconstructed information, we can formalize the unlearned data perturbation as follows to find the suitable perturbation. 
\begin{equation} \label{perturb_loss}
	\small
	\begin{aligned}
			&\min_{\Delta^p} \mathcal{L}_{\texttt{AE}} (\hat{X_u}', X_u + \Delta^p) \\
			&\text{s.t.} \hspace{3mm}  \Delta^p \in \arg \min_{\theta_{ \backslash (X_u + \Delta^p)}} \sum_{x \in D_r} \ell (x;\theta_{\backslash (X_u + \Delta^p)})
	\end{aligned}
\vspace{-2mm}
\end{equation}
where $\hat{X_u}' = \texttt{AE}(\delta_{\backslash (X_u + \Delta^p)})$, meaning that the samples are reconstructed based on the posterior difference that unlearns the perturbed data $X_u + \Delta^p$. We define the constraint that $ \Delta^p : \| \Delta^p \|_{\infty} \leq \alpha$ to ensure that the perturbed data will not be too different from the original data. We can combine these two losses together and treat them as two objectives, thus can be optimized with two-objective optimization methods \cite{sener2018multi,poirion2017descent,guo2020learning}. During the perturbation optimization process, we fix the trained model $\theta_t$ and the reconstruction model $\texttt{AE}$. We only update the perturbation $\Delta^p$ of $X_u$ to induce an augmented unlearned posterior difference $\delta_{\backslash (X_u + \Delta^p)}$, which improves the reconstruction effect. To find an effective perturbation, we can employ the restars technique, which is inspired from \cite{GeipingFHCT0G21,QinMGKDFDSK19}, and we provide the corresponding algorithm in \Cref{UDP_algorithm}.

\noindent
\textbf{Unlearning Influence-based Division after Unlearning.} 
The unlearning influence-based division strategy utilizes the convenient properties of the first-order data influence estimation. After achieving the overall posterior differences for multiple samples $\delta_{\backslash D_u}$, the user can quickly estimate the basic data influence for each integrated sample $x_u \in D_u$, and we divide the overall posterior difference according to the weight of each sample's influence. We assume the divided posterior difference of the integrated sample obeys a Gaussian distribution:
\begin{equation} 
	\vspace{-2mm}
	\small
	\begin{aligned}
		\delta_{\backslash x_u}  \sim \mathcal{N}(\frac{\delta_{\backslash D_u} }{ \sum_{x_u \in D_u} \nabla  \ell (x_u;\theta_t)} \cdot \nabla  \ell (x_u;\theta_t), \sigma^2), \\
		\text{s.t.} \   \delta_{\backslash D_u} = \sum_{x_u \in D_u}  \delta_{\backslash x_u} ,
	\end{aligned}
\end{equation}
where we keep the divided posterior difference values as the mean and add a random deviation to it; meanwhile, we keep the sum of all the split slice posterior differences $\sum_{x_u \in D_u}  \delta_{\backslash x_u}$ equal to the overall posterior difference $\delta_{\backslash D_u}$. Thus, we ensure every reconstruction has a unique divided posterior difference without additional change or noise in the original $\delta_{\backslash D_u}$. This operation changes the reconstruction task for multiple samples based on the same $\delta_{\backslash D_u}$ as reconstructing every single sample of $D_u$ based on multiple divided posterior differences.

\section{Performance Evaluation}
\subsection{Settings} \label{exp_setting}




\noindent
\textbf{Datasets.}
We conducted experiments on four widely adopted public datasets: MNIST \cite{deng2012mnist}, CIFAR10 \cite{krizhevsky2009learning}, STL-10 \cite{coates2011analysis}, and CelebA \cite{liu2018large}. These datasets offer a range of objective categories with varying levels of learning complexity. 

\noindent
\textbf{Models.}
In our experiments, we select a 5-layer multi-layer perceptron~(MLP) connected by ReLU, a 7-layer convolutional neural network (CNN), and ResNet-18. Specifically, we use two 5-layer MLP models, one as the encoder and one as the decoder, to consist of the reconstructor for MNIST. We use two 7-layer CNNs to consist of the reconstructor for CIFAR10, STL-10, and CelebA. The main structure of the ML service model is implemented with a ResNet-18. Moreover, to align with existing backdoor-based verification methods, we also train a Verifier model using a 5-layer MLP model, which is trained after reconstruction. The Verifier model training algorithm is presented in \Cref{verifi_train}.




\noindent
\textbf{Metric.}
We use three metrics, model accuracy, reconstruction similarity, and verifiability, to measure the ability previously defined for the unlearning auditing scheme. Moreover, we use the running time to assess the methods' efficiency. We briefly summarize the metric as follows. 
\begin{itemize}[itemsep=0pt, parsep=0pt, leftmargin=*]
	\item \textbf{Accuracy.} Model accuracy evaluates functionality preservation and shows whether the auditing methods influence the utility of the service model. 
	\item  \textbf{Reconstruction Similarity.} It evaluates how much information about the specified samples is unlearned by reconstructed cosine similarity, as introduced in \Cref{similarity_eq}.
	\item \textbf{Verifiability.} Verifiability is used to measure the data removal verification by calculating the correct classifying rate of the $\text{Verifier}$, which is defined in \Cref{verifiability_def} in \Cref{detailed_metrics}. 
	\item  \textbf{Running Time.} It is used to assess the efficiency, which records the running time of the entire process of each method.
\end{itemize}


\noindent
\textbf{Compared Unlearning Verification Benchmarks.}
There are mainly three data removal verification solutions \cite{hu2022membership,sommer2022athena,guo2023verifying}, all based on backdooring methods. 
We only compare our method with MIB \cite{hu2022membership} because MIB is the most popular and has the best verification effect among these three methods. Note that since these methods can only support verifying the backdoored samples, most of the evaluation for MIB is verifying for unlearning only backdoored samples; our method is verifying for unlearning genuine samples.


\noindent
\textbf{Unlearning Benchmarks.} 
The evaluation for unlearning verification methods is conducted on four mainstream unlearning algorithms: SISA~\cite{bourtoule2021machine}, HBU~\cite{guo2019certified}, VBU~\cite{nguyen2020variational} and RFU~\cite{wang2023machine}.

\begin{figure}[t]
	\centering
	\includegraphics[width=0.99\linewidth]{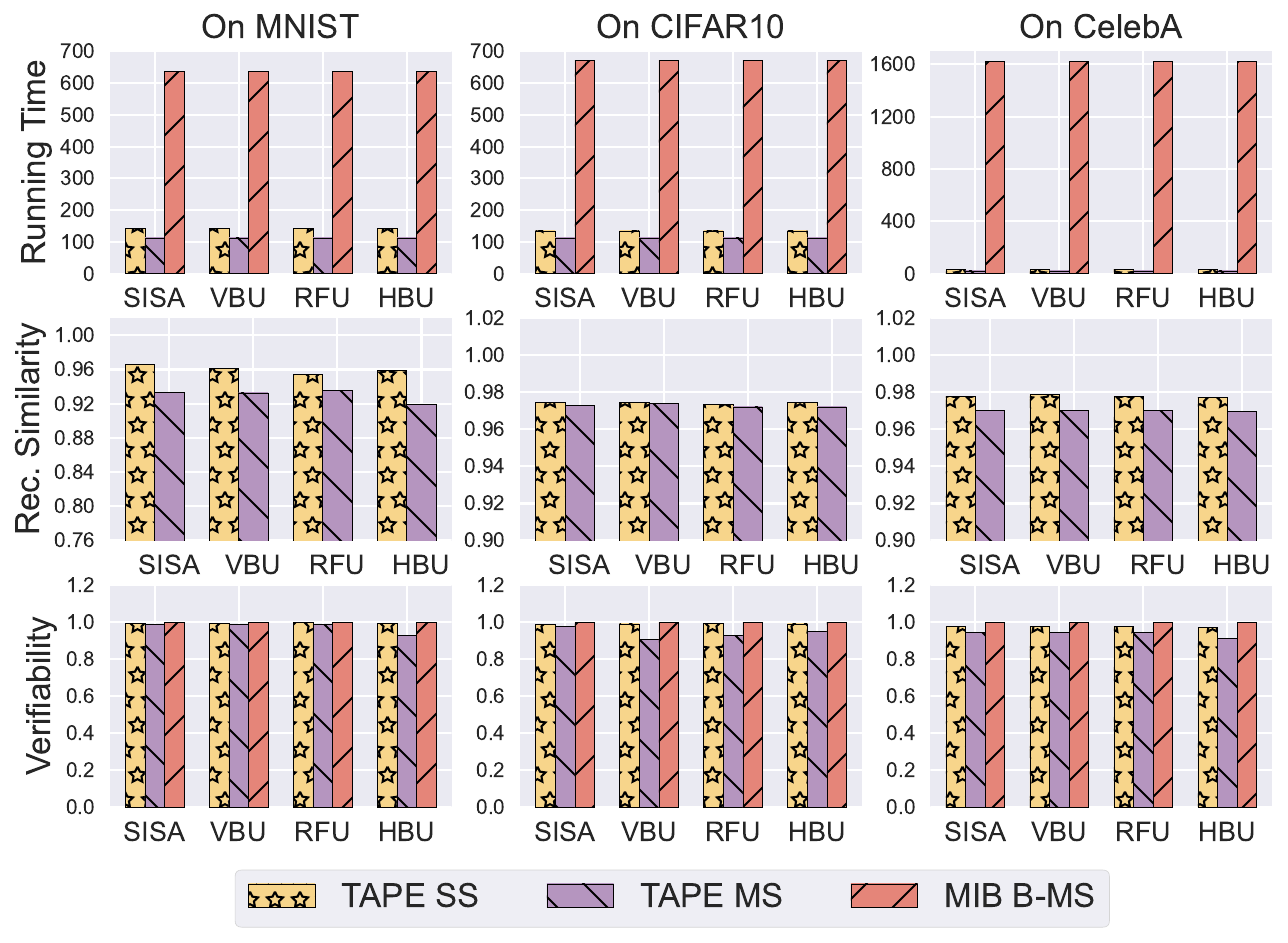}
	\vspace{-2mm}
	\caption{Auditing for different unlearning methods. TAPE consistently achieves significant efficiency improvement and a better unlearning auditing effect for a single sample (SS) than for multiple samples (MS). 
		\vspace{-5mm}
	} 
	\label{evaluation_of_unl}
\end{figure}

\subsection{Evaluations of Unlearning Auditing based on Various Unlearning Benchmarks} \label{exp_on_unl_benchmarks}
 
 
\noindent
\textbf{Setup.} 
We demonstrate the evaluation of unlearning auditing methods on four mainstream unlearning benchmarks in \Cref{evaluation_of_unl}.
We evaluate unlearning scenarios of both single-sample (SS) where the Erased Sample Size ({\it ESS}) is 1 and multi-sample (MS) where {\it ESS}=20. Since the MIB method is unable to verify only for unlearning genuine samples, we here evaluate the verification of MIB for backdoored multi-samples (B-MS), $D_{b} \gets (X_b + \text{trigger}, Y_{target})$, which add a white block patch as the trigger at the right bottom of chosen images and change the corresponding labels for the backdooring target. When evaluating TAPE, to keep the setting similar to MIB, we add the perturbation with the same limit distance as the trigger patch to the genuine unlearned samples but do not change the labels for backdooring, $D_{u,p} \gets (X_u+\Delta^p, Y_{u})$, which is achieved through the unlearned data perturbation (UDP) method.


\noindent
\textbf{Evaluations of Auditing Efficiency.} 
The first row of \Cref{evaluation_of_unl} shows the running time of TAPE and MIB across different unlearning algorithms on MNIST, CIFAR-10, and CelebA. TAPE significantly outperforms MIB in terms of running time, as TAPE only involves the unlearning training process. The greatest speedup is observed on CelebA. Additionally, TAPE takes more time for single-sample unlearning auditing compared to multi-sample unlearning auditing, as training the reconstructor model on a single-sample level for a local dataset requires more time.

\noindent
\textbf{Evaluations of Unlearning Auditing Effect.} 
The second row, which depicts reconstruction similarity, illustrates the evaluation results of how much information about the specified samples has been unlearned. As MIB is unable to measure the extent of information unlearned from the model, it is omitted in this row. Among all unlearning methods (SISA, VBU, RFU, and HBU), TAPE achieves better reconstruction similarity for single-sample unlearning than for multi-sample unlearning. This suggests that unlearning a single sample tends to reveal more information about the erased sample in the unlearning posterior difference.

The third row shows the comparison with MIB by the verifiability of data removal verification. All methods have a high verifiability result. It indicates that all unlearning methods are effective in these evaluations to answer if the samples are unlearned from the model. However, we should note that in the experiments, MIB only verifies the backdoored samples $D_{u,b}$, while TAPE can verify the genuine samples $D_{u,p}$, which has kept the original labels.

We also demonstrate an overall evaluation for MIB and TAPE on SISA \cite{bourtoule2021machine} in \Cref{eval_on_C_and_S}. Here, we evaluate auditing genuine samples for both MIB and TAPE instead of setting backdoored samples for MIB. TAPE achieves effective auditing results as analyzed in \Cref{evaluation_of_unl}. However, the MIB cannot successfully verify the unlearning of any genuine samples in the Unl. Verifiability of \Cref{eval_on_C_and_S}. Moreover, since the TAPE scheme is independent of the original model training process, it will not influence the model utility of the original ML service model, keeping the same model accuracy as the ``Original''.  We present additional experimental results in \Cref{overall_eval_app}.

\begin{figure}[t]
	\centering
	\vspace{-2mm}
	\hspace{-6mm}
	\subfloat{ 	\label{fig:mnistbackaccbetacurve} 
		\includegraphics[scale=0.205]{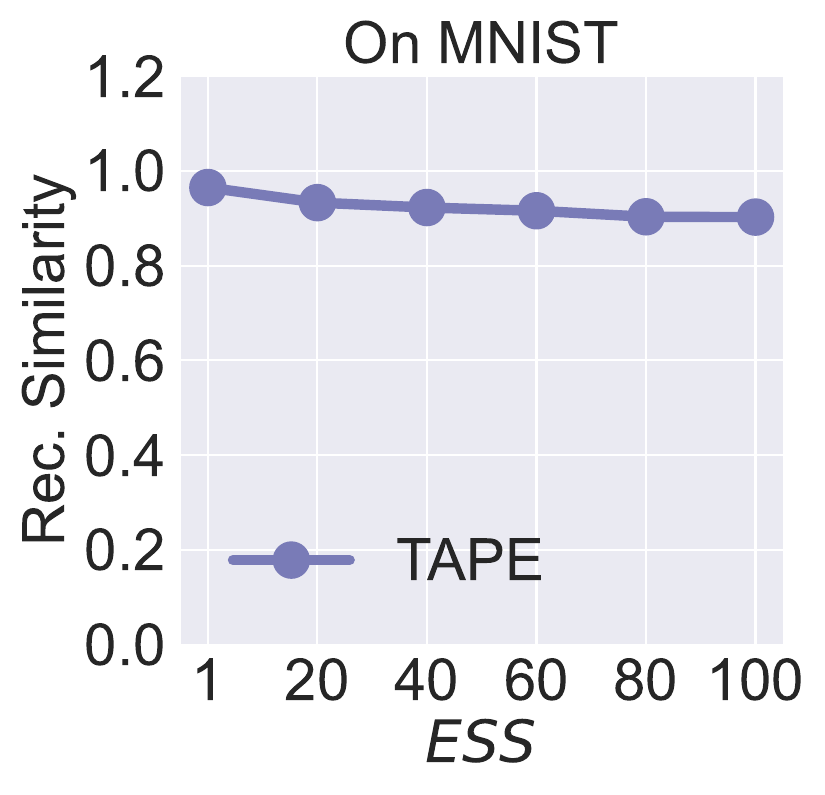}
	}		
	\hspace{-3mm}
	\subfloat{  	\label{fig:cifar10recsimsamplesizesimanalysis}
		\includegraphics[scale=0.205]{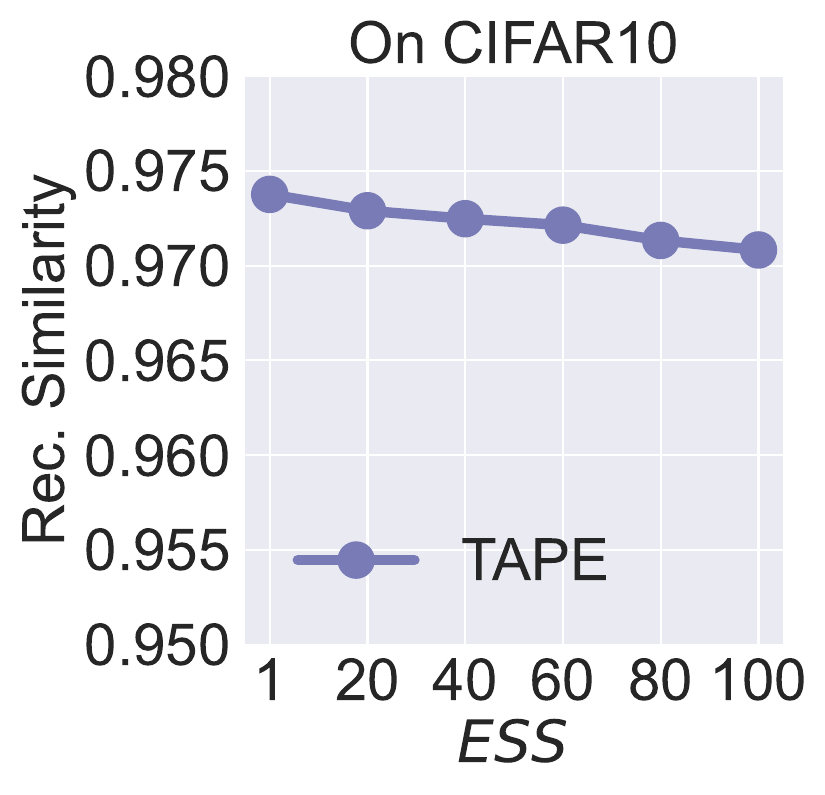}
	}	
	\hspace{-3mm}
	\subfloat{ \label{fig:celebarecsimsamplesizesimanalysis}
		\includegraphics[scale=0.205]{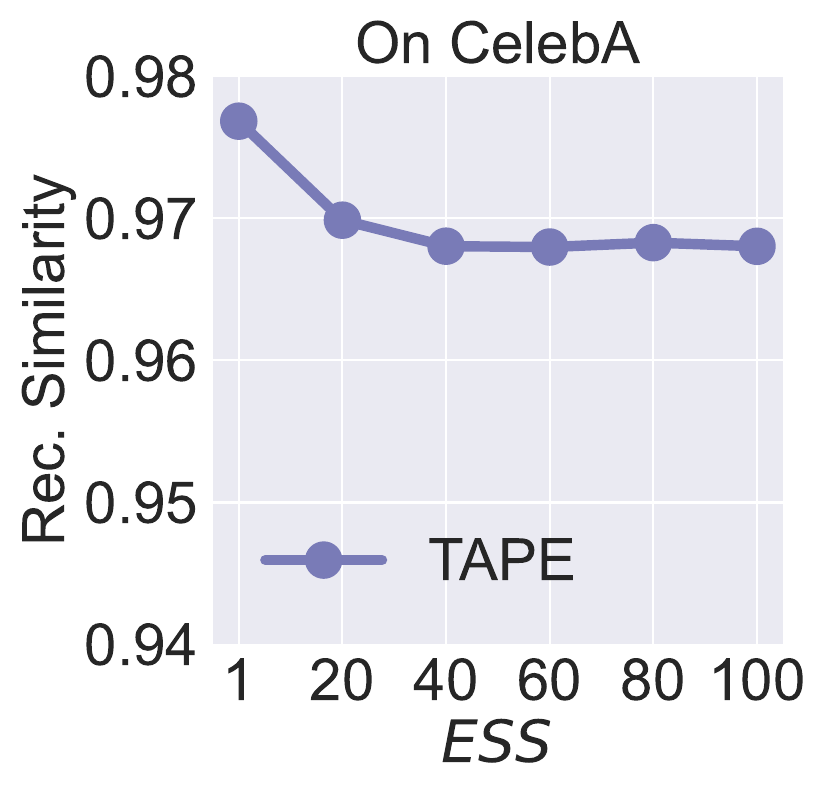}
	} \vspace{-2mm}
	\\
	\hspace{-6mm}
	\subfloat{ \label{fig:mnistaccbetacurve}
		\includegraphics[scale=0.205]{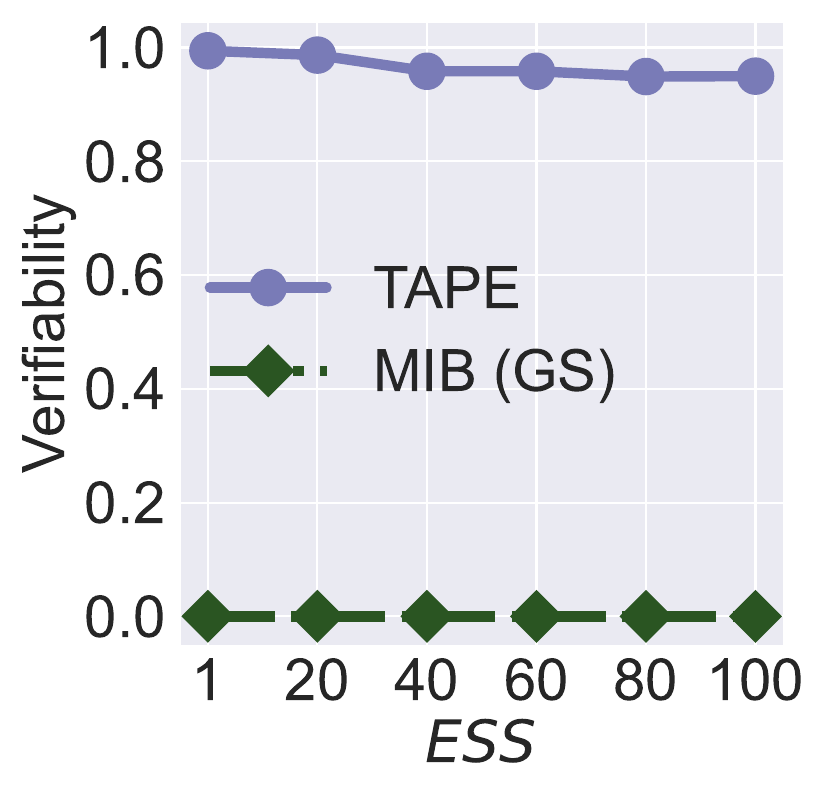}
	}
	\hspace{-3mm}
	\subfloat{ \label{fig:cifar10verifiabilitysamplesizesimanalysis}
		\includegraphics[scale=0.205]{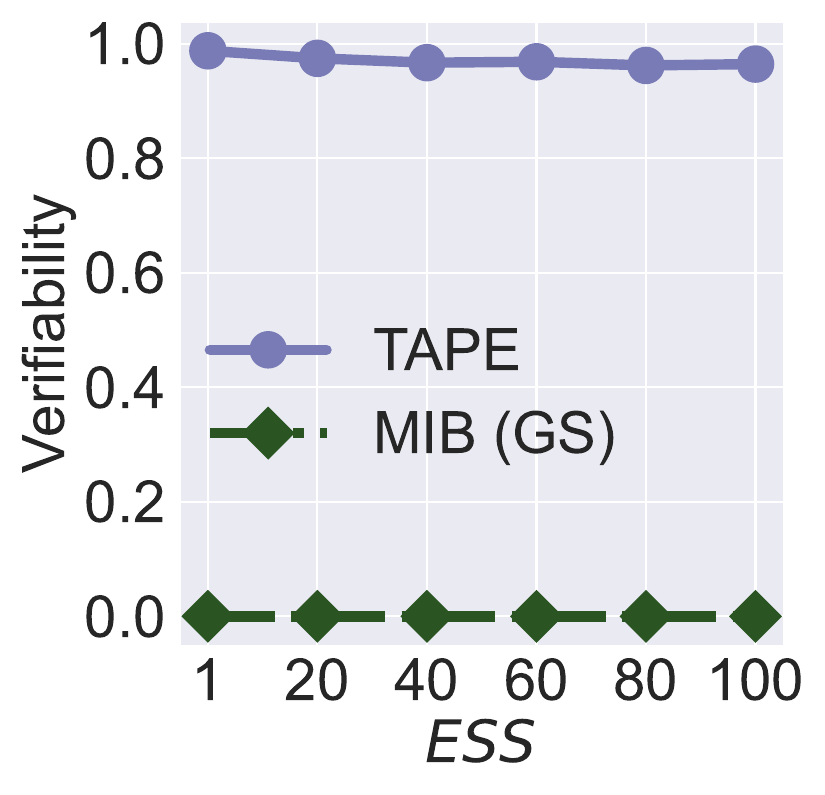}
	}
	\hspace{-3mm}
	\subfloat{ 	\label{fig:celebaverifiabilitysamplesizesimanalysis}
		\includegraphics[scale=0.205]{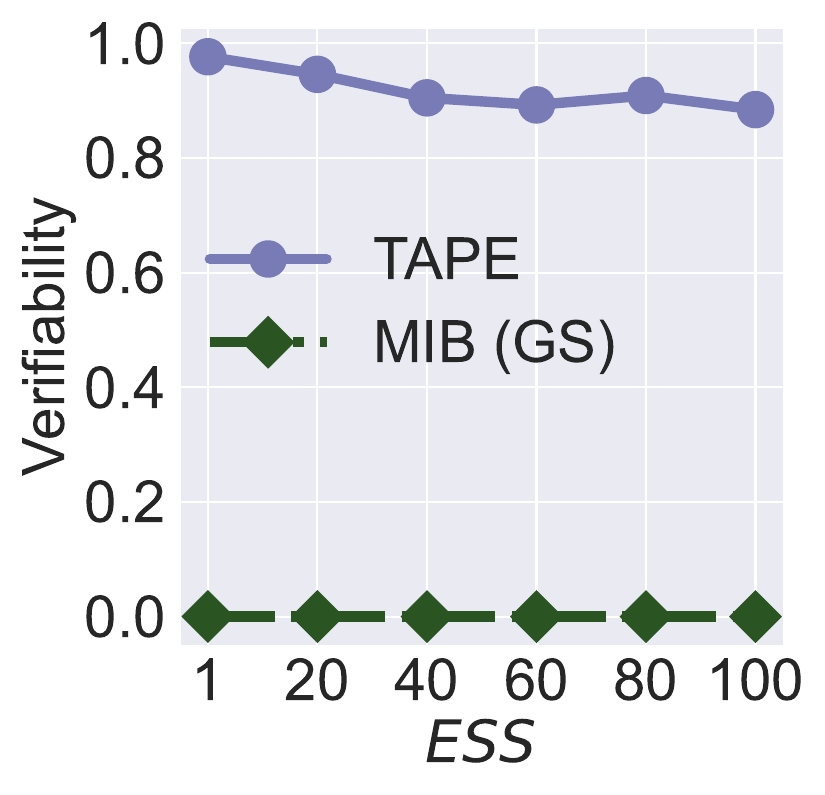}
	}  \vspace{-2mm}
	\caption{Evaluations of impact about different $\text{\it ESS}$. Here, we evaluate the unlearning verification of genuine samples (GS) rather than backdoored samples for MIB.  \vspace{-2mm}} 
	\label{evaluation_of_in_or_not_in} 
\end{figure}

 \begin{table}[t]
	\scriptsize
	\caption{Evaluation Results on CIFAR10 and STL-10\vspace{-2mm}}
	\label{eval_on_C_and_S}
	\resizebox{\linewidth}{!}{
		\setlength\tabcolsep{2.pt}
		\begin{tabular}{c|cccccc}
			\toprule[1pt]
			\multirow{2}{*}{} & \multicolumn{3}{c} {CIFAR10, $\text{\it ESS}=20$} & \multicolumn{3}{c} {STL-10, $\text{\it ESS}=2$} \\
			\cmidrule(r){2-4}   \cmidrule(r){5-7}
			& Original & MIB \cite{hu2022membership}   	 & TAPE & Original & MIB   & TAPE\\
			\midrule 
			Running time (s)   & 644   & 673 			& \textbf{113} 	 & 781	& 809  & \textbf{74.90}    \\
			Model Acc.	         & \textbf{81.62\%}    & 79.13\%    & \textbf{81.62\%} & \textbf{68.99\%} &  67.26\% &  \textbf{68.99\%}  \\
			Rec. Sim.  			 	&-& -									& \textbf{0.973}					 	&- & - & \textbf{0.174}  \\
			Unl. Verifiability     & $0.00\%$ & $0.00\%$ & \textbf{97.44\%}  	& $0.00\%$ & $0.00\%$ & \textbf{84.40\%}\\
			\bottomrule[1pt]
	\end{tabular}}
		\vspace{-4mm}
\end{table}

 \begin{figure*}[t]
	\centering
	\includegraphics[width=0.9\linewidth]{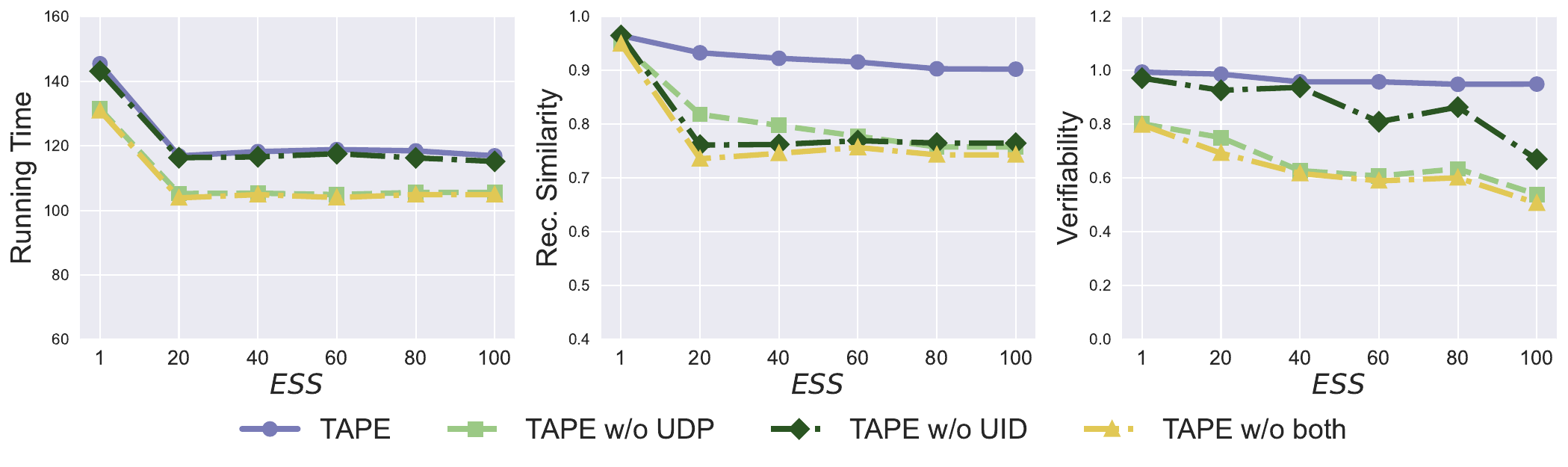}
	\vspace{-4mm}
	\caption{Ablation study about the unlearned data perturbation (UDP) and unlearning influence-based division (UID) strategies of TAPE on MNIST. The legends stand for the entire TAPE, TAPE without (w/o) the UDP strategy, TAPE w/o the UID strategy, and TAPE w/o both strategies.}
	\vspace{-4mm}
	\label{fig:ablationexp}
\end{figure*}

\subsection{Ablation Study of Erased Samples Size ($\text{\it ESS}$)} \label{impact_of_ess}
\noindent
\textbf{Setup.}
\Cref{evaluation_of_in_or_not_in} illustrates the impact of $\text{\it ESS}$ when providing unlearning auditing on MNIST, CIFAR10 and CelebA. The largest $\text{\it ESS}$ is 100 in this experiment, which is around $0.2\%$ of training datasets in MNIST and CIFAR10. In practice, $0.2\%$ data might be very large for unlearning, which has been analyzed in \cite{bertram2019five,chen2021machine}. To illustrate a better comparison, we verify the unlearning of genuine perturbed samples $D_{u,p}$ for both MIB and TAPE. The ablation study is conducted based on another representative unlearning method, VBU \cite{nguyen2020variational}, to demonstrate the scalability of our method for different unlearning methods. Due to the page limitation, we present the impact on the efficiency of $\text{\it ESS}$ in \Cref{imp_of_efficiency_ESS}.
 

\noindent
\textbf{Impact on Unlearning Auditing.}
\Cref{evaluation_of_in_or_not_in} shows the evaluation of both auditing how much information is unlearned (the first row) and data removal status verification (the second row) on MNIST, CIFAR10 and CelebA. When $\text{\it ESS}=1$, TAPE can effectively provide the auditing of unlearning information and data removal status on all datasets. By contrast, MIB cannot support the data removal verification of genuine samples (black dotted line in the second row in \Cref{evaluation_of_in_or_not_in}). Moreover, both two evaluation metrics have a decreasing trend when $\text{\it ESS}$ increases.

 \subsection{Ablation Study of Two Strategies}
 

 \noindent
 \textbf{Setup.}
 We conduct the experiments on MNIST in four situations. ``TAPE'' means the entire scheme with the two strategies. ``TAPE w/o UDP'' means TAPE without the UDP strategy while keeping the UID strategy, and ``TAPE w/o UID'' means that we remove the UID strategy of TAPE while keeping the UDP strategy. The ``TAPE w/o both'' means we remove both strategies of TAPE for auditing.

 \noindent
 \textbf{Impact on Efficiency.} We present the efficiency evaluation in the first sub-figure of \Cref{fig:ablationexp}. Since the UDP strategy introduces $R=10$ restarts training to find perturbations for erased samples, it consumes around 10 seconds in TAPE. Compared with UDP, UID almost does not consume computational time, as the main time consumption is the unlearned shadow model building and reconstructor training.

 \noindent
 \textbf{Impact on Auditing Unlearning Effectiveness.} The second sub-figure in \Cref{fig:ablationexp} shows the reconstruction similarity of different methods. All methods achieve a high reconstruction similarity when $\it{ESS} = 1$, which shows the effectiveness of TAPE in single-sample unlearning auditing even without the two strategies. However, when the posterior difference contains information of multiple samples, $\it{ESS} > 1$, if we don't have the UID strategy, the reconstruction similarity drops dramatically, showing as ``TAPE w/o UID'' and ``TAPE w/o both''. The UID plays a vital role in the reconstruction of multiple samples. While the UDP strategy also enhances the reconstruction quality, as shown in ``TAPE w/o UDP'',  its impact is not as substantial as the UID strategy. By contrast, in the third sub-figure in \Cref{fig:ablationexp}, the UDP strategy impacts the verifiability of data removal than the UID strategy, showing as ``TAPE w/o UDP'' and ``TAPE w/o UID''. These results show the significant improvement of the two strategies in benefiting unlearning auditing of how much information is unlearned and data removal status.

 \begin{figure*}[t]
 	\centering
 	\hspace{-3mm}
 	\subfloat{  	\label{fig:mnistrunningtimenoiseanalysisbar}
 		\includegraphics[scale=0.25]{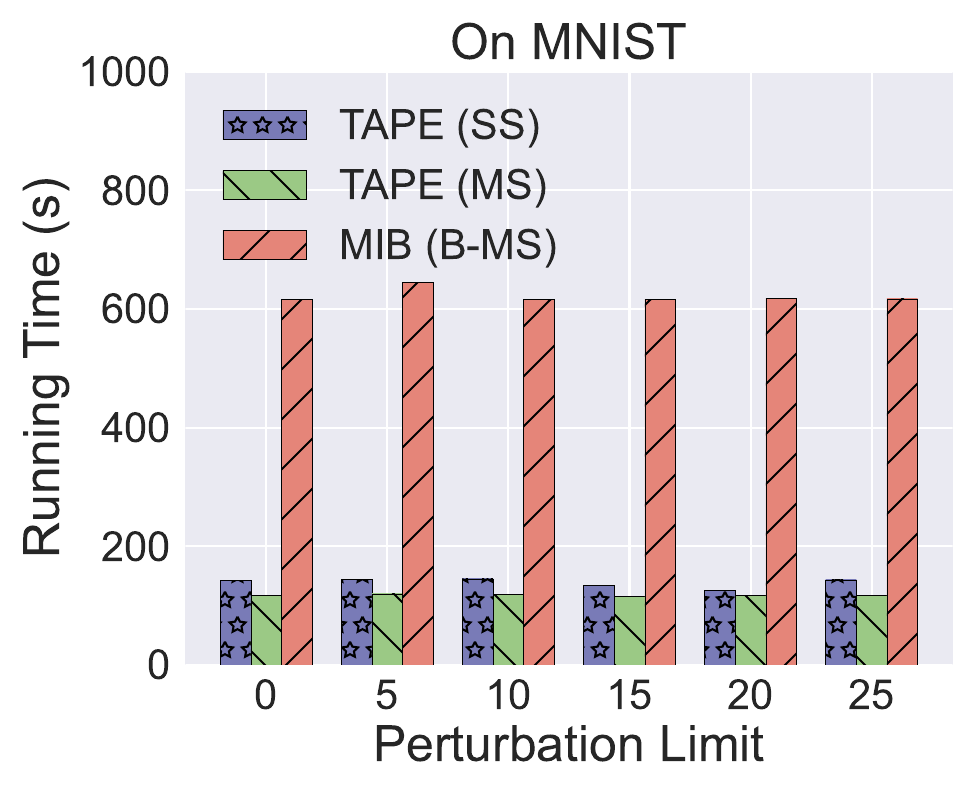}
 	} 
 	\hspace{-2mm}
 	\subfloat{  	 	\label{fig:cifar10runningtimenoiseanalysisbar}
 		\includegraphics[scale=0.25]{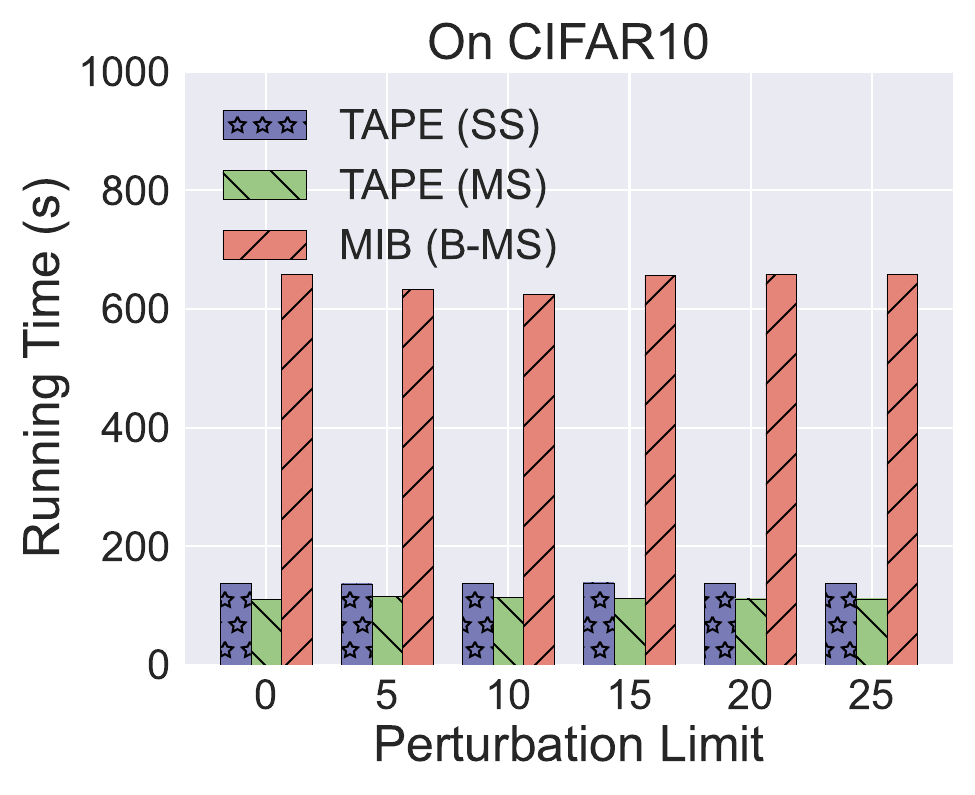}
 	}
 	\hspace{-2mm}
 	\subfloat{ 		\label{fig:stl10rtnoiseanalysisbar}
 		\includegraphics[scale=0.25]{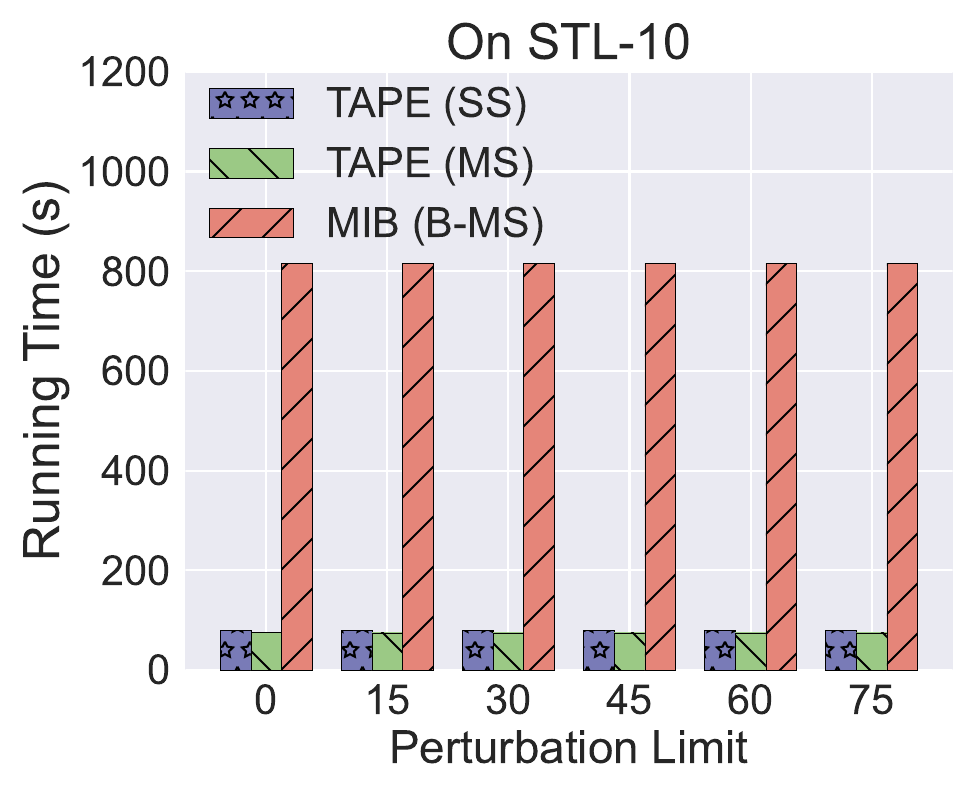}
 	}			
 	\hspace{-2mm}
 	\subfloat{ 	 	\label{fig:celebarunningtimenoiseanalysisbar}
 		\includegraphics[scale=0.25]{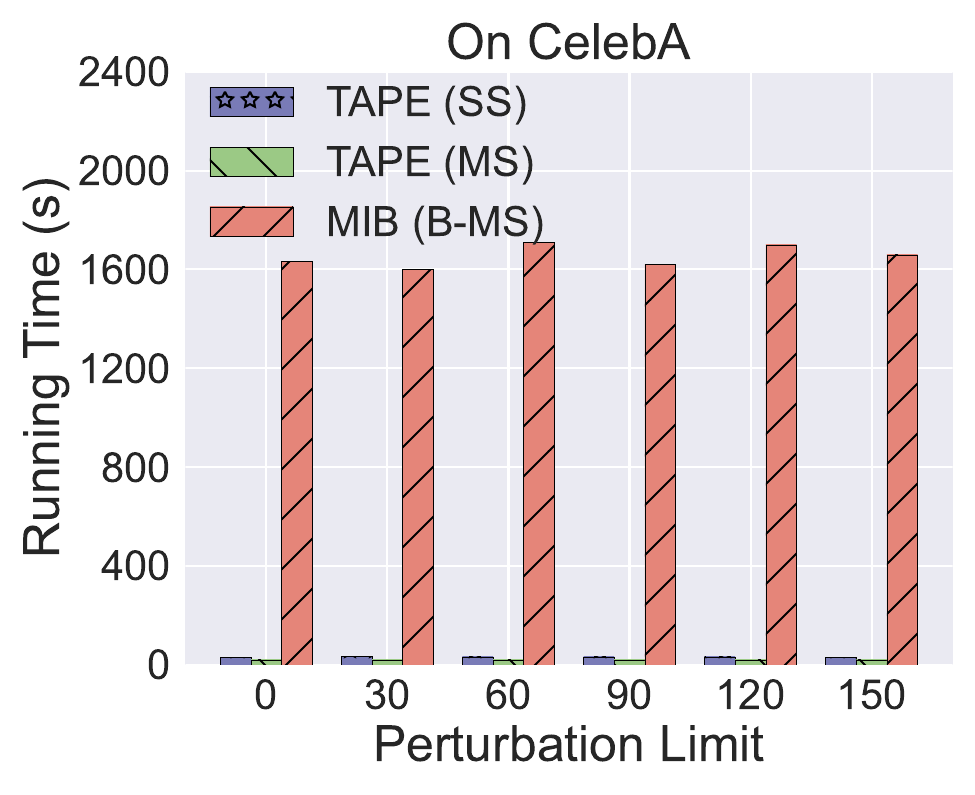}
 	} \vspace{-3mm} \\
 	\hspace{-3mm}
 	\subfloat{ 	  	\label{fig:mnistrecmsenoiseanalysis}
 		\includegraphics[scale=0.25]{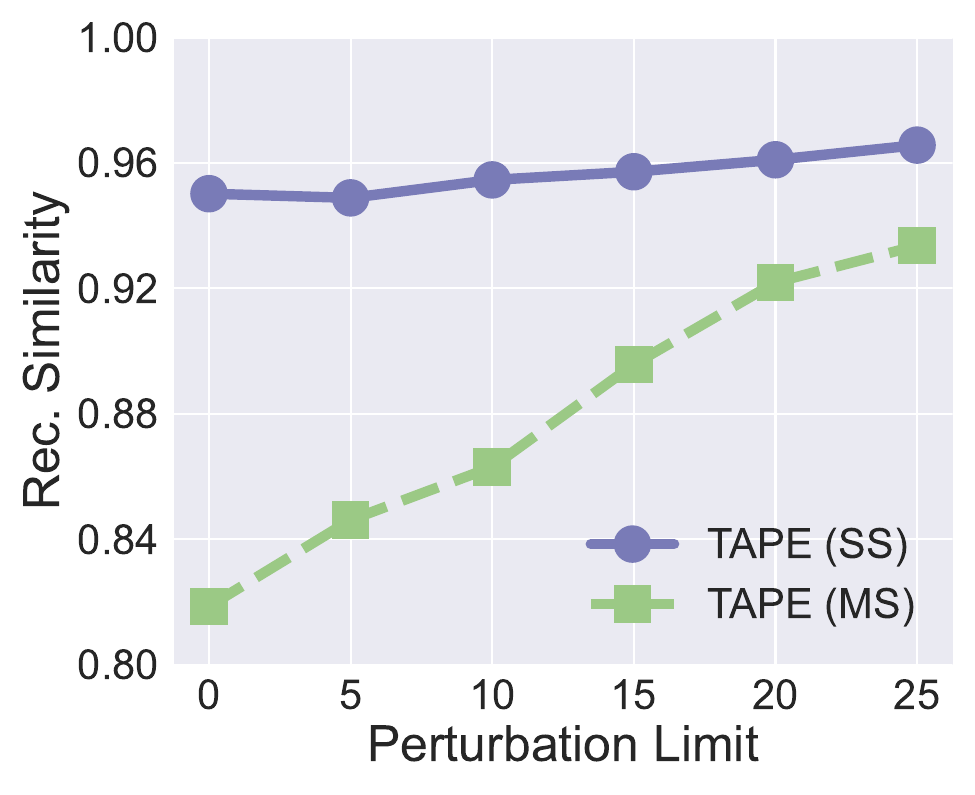}
 	}			
 	\hspace{-2mm}
 	\subfloat{   	\label{fig:cifar10recmsenoiseanalysis}
 		\includegraphics[scale=0.25]{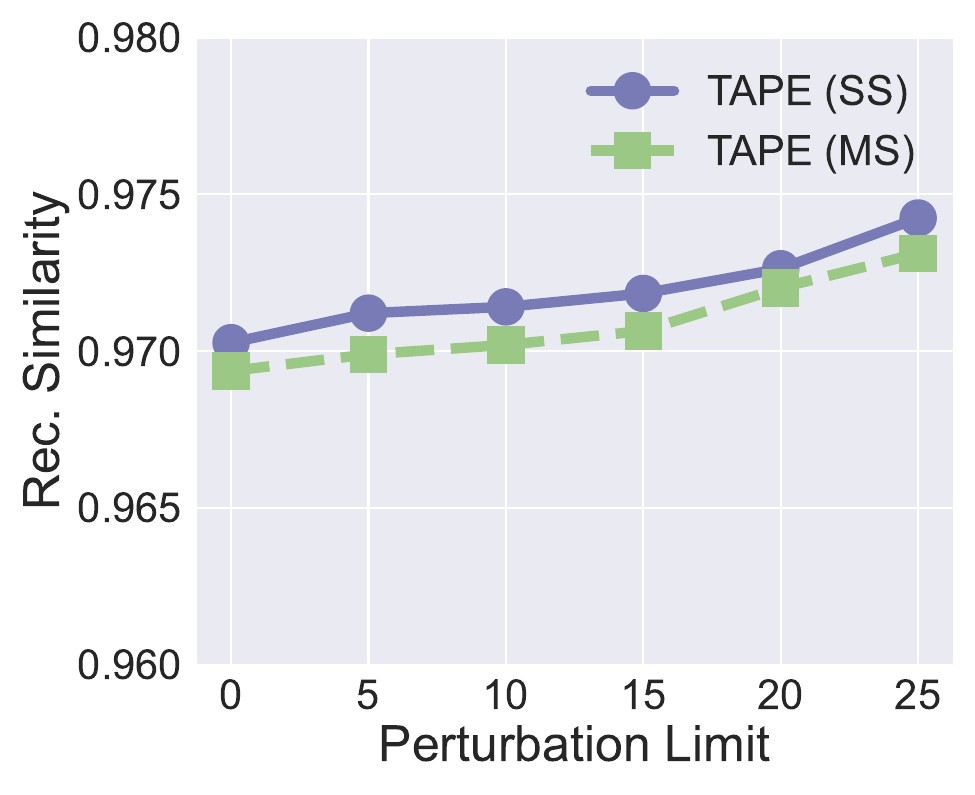}
 	} 
 	\hspace{-2mm}
 	\subfloat{     	\label{fig:stl10recmsenoiseanalysis}
 		\includegraphics[scale=0.25]{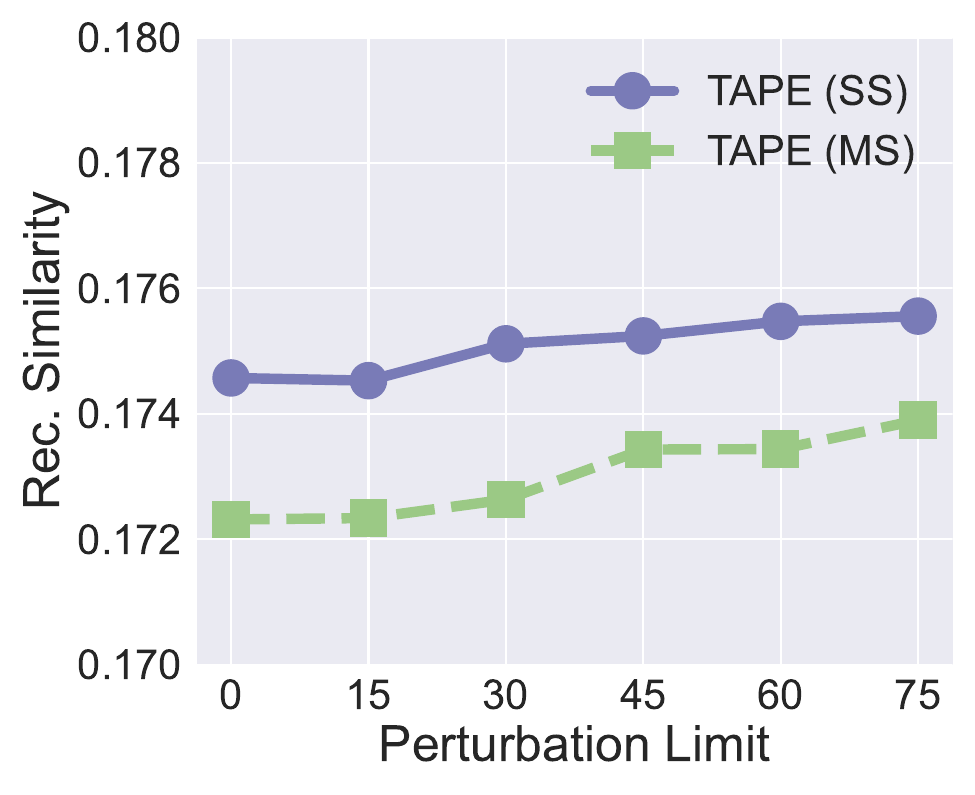}
 	}
 	\hspace{-2mm}
 	\subfloat{  	\label{fig:celebarecmsenoiseanalysis}
 		\includegraphics[scale=0.25]{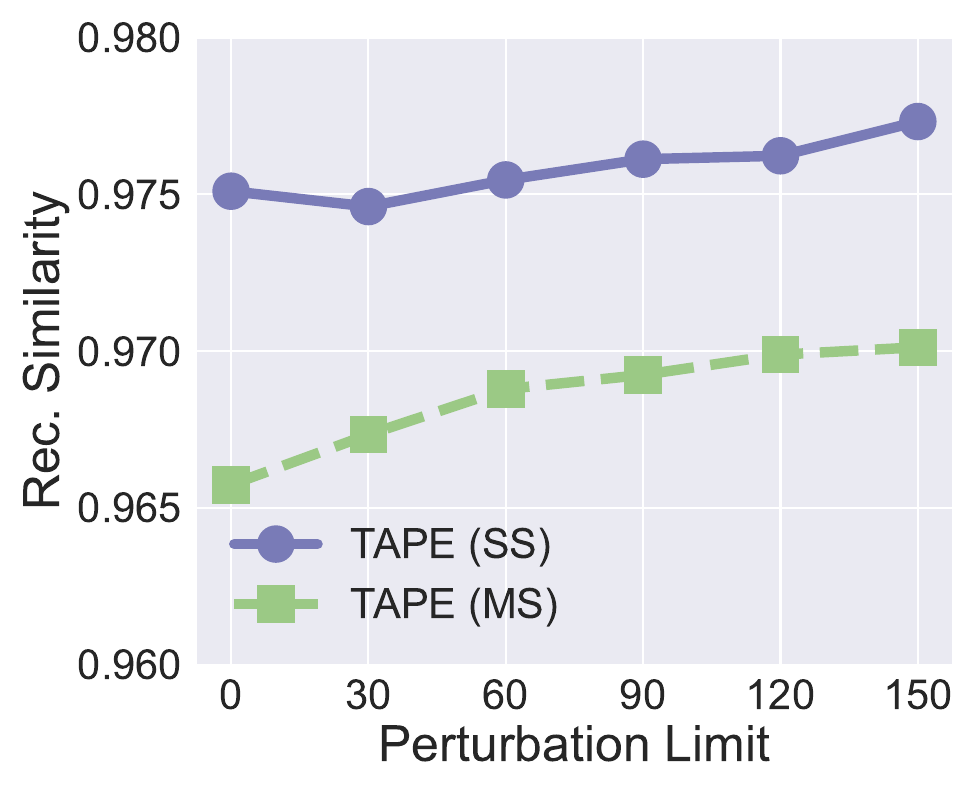}
 	} \vspace{-3mm}  \\
 	\hspace{-3mm}
 	\subfloat{ 	\label{fig:mnistverifiabilitynoiseanalysis}
 		\includegraphics[scale=0.25]{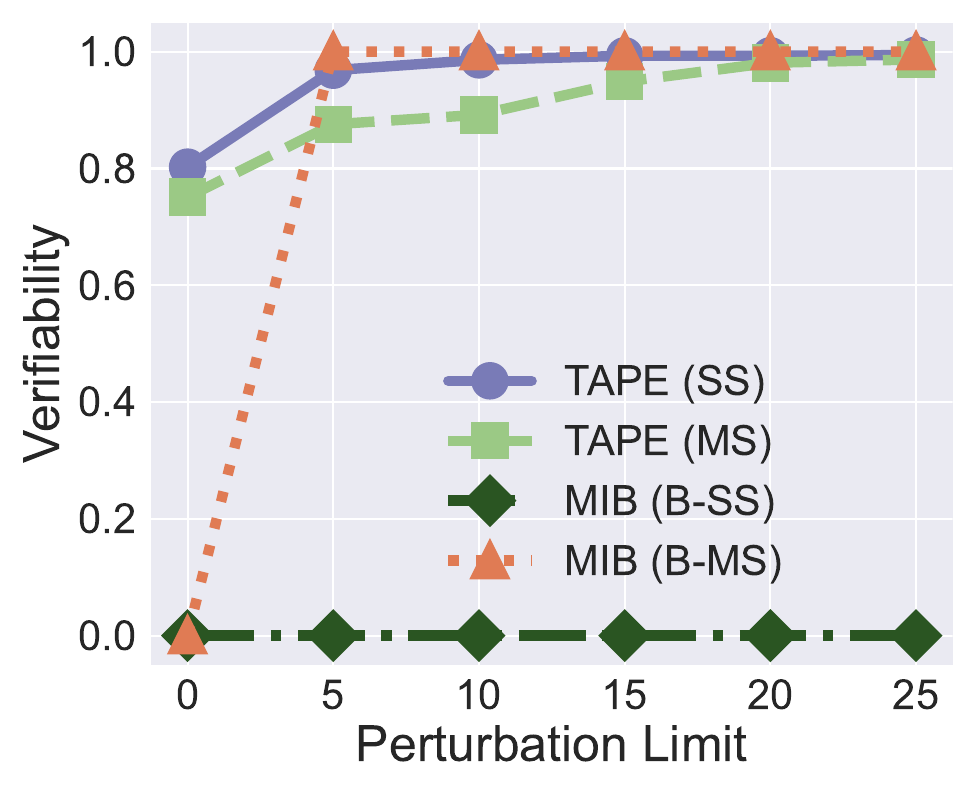}
 	}			
 	\hspace{-2mm}
 	\subfloat{   	\label{fig:cifar10verifiabilitynoiseanalysis}
 		\includegraphics[scale=0.25]{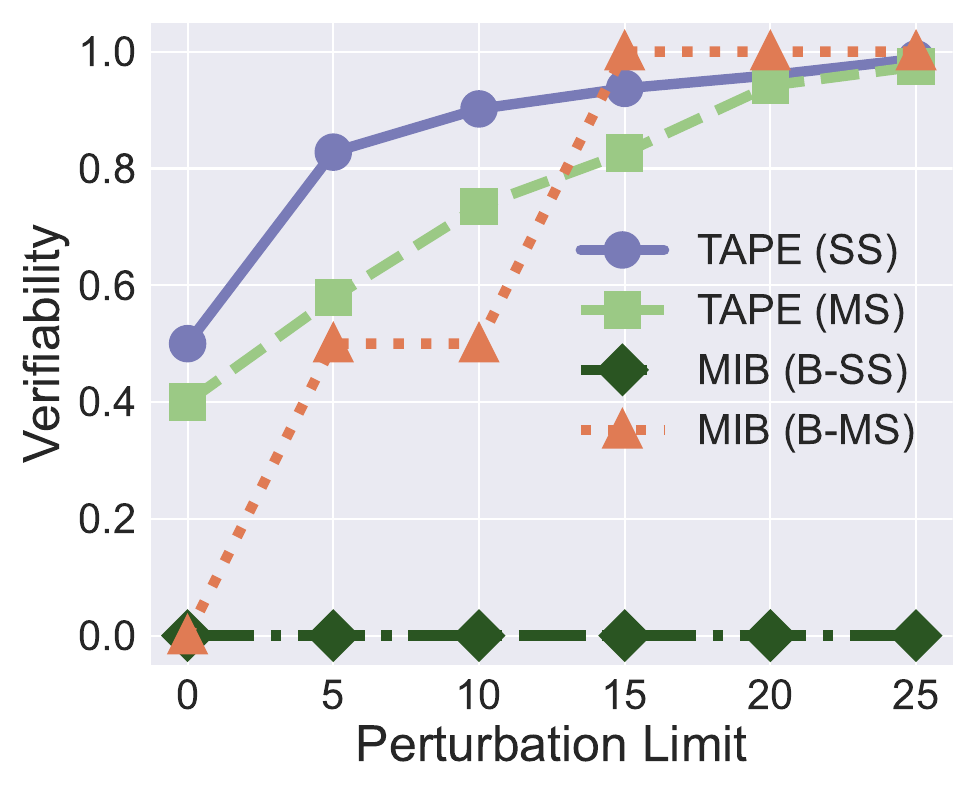}
 	} 
 	\hspace{-2mm}
 	\subfloat{  	\label{fig:stl10verifiabilitynoiseanalysis}
 		\includegraphics[scale=0.25]{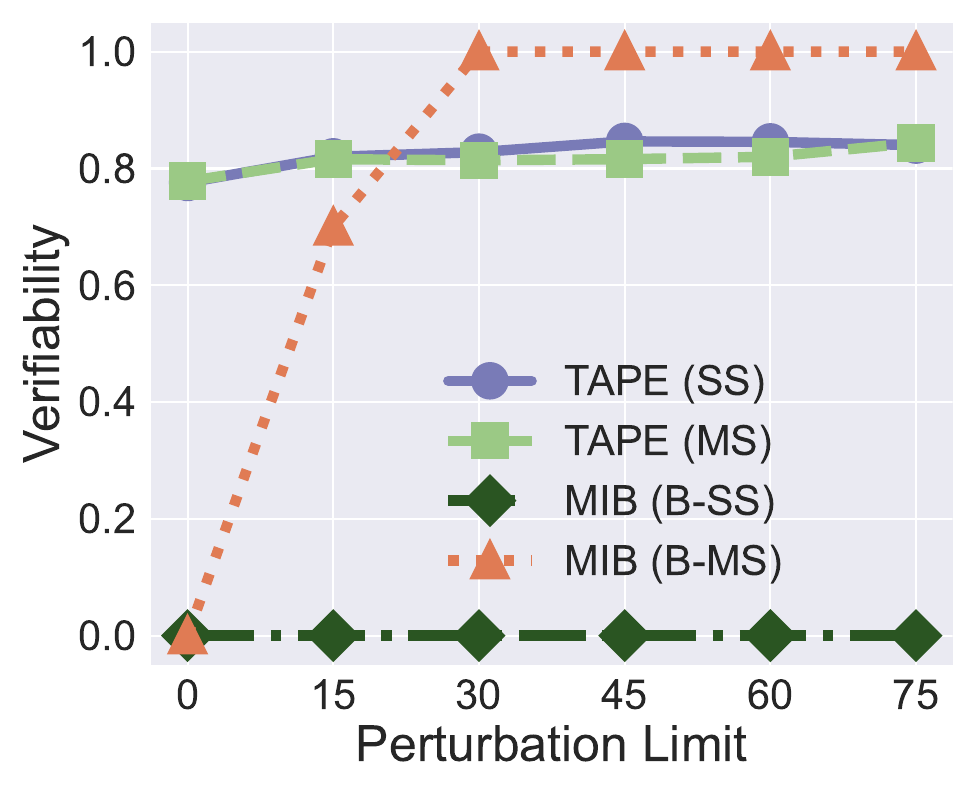}
 	}
 	\hspace{-2mm}
 	\subfloat{   	\label{fig:celebaverifiabilitynoiseanalysis}
 		\includegraphics[scale=0.25]{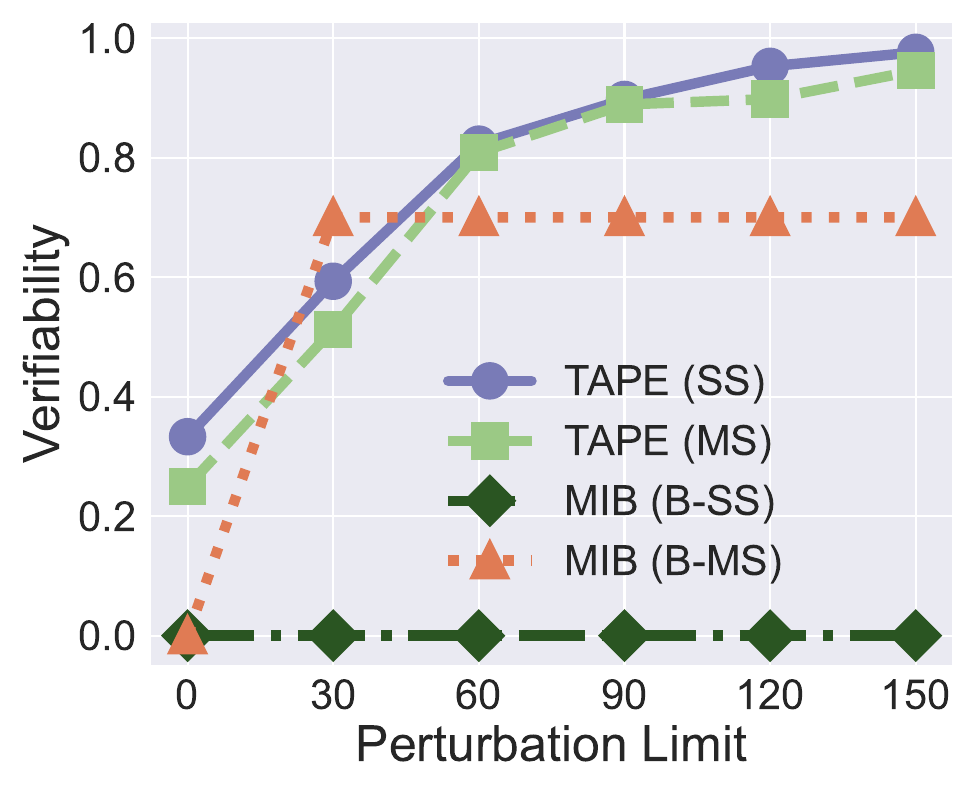}
 	}
 	\caption{Evaluations of the impact of the unlearned data perturbation limit. ``SS'' stands for single-sample unlearning scenario, and ``MS'' means multi-sample unlearning scenario. ``B-SS'' means backdoored single-sample scenario, and ``B-MS'' means backdoored multi-sample unlearning scenario.} 
 	\vspace{-2mm}
 	\label{evaluation_of_noise} 
 \end{figure*}

\subsection{Detailed Ablation Study of UDP} \label{adding_noise}

We additionally evaluate the impact of unlearned data perturbation. We find that only perturbation without changing the labels as backdoor-based methods already significantly improves the information reconstruction and assists the verifiability of data removal.

\noindent
\textbf{Setup.} 
In this experiment, we keep all other parameters fixed while only changing the perturbation limitation value $\alpha$. As introduced in \Cref{two_s}, the UDP outputs $D_{u,p} \gets (X_u + \Delta^p, Y_u)$, and the perturbation is limited as $\| \Delta^p \|_{\infty} \leq \alpha$. We set the perturbation limit distance value from 0 to 25 on MNIST and CIFAR10, 0 to 75 on STL-10, and 0 to 150 on CelebA, which is determined by the data size. We only perturb the unlearned data but do not change the corresponding labels to ensure the utility of the genuine samples. To better illustrate the impact of $\alpha$, we keep a copy of original data $D_u$ in the remaining dataset, which is the hardest scenario for unlearning effectiveness auditing. 
The unlearning method employed here is the approximate unlearning method VBU. 
We evaluate our method in both single-sample (SS) and multi-sample (MS) unlearning requests. For the MIB method, we evaluate in the backdoored single-sample (B-SS) scenario, and we control the backdooring trigger patches with the sample distance limitation value as our methods to ensure they can be compared. The experimental results on MNIST, CIFAR10, STL-10, and CelebA are presented in \Cref{evaluation_of_noise}.

 \noindent
 \textbf{Impact on Efficiency.}
The first column illustrates the running time of TAPE and MIB, which clearly demonstrates the improvement of TAPE in terms of efficiency. The reason is that the verification models of TAPE are trained independently of the original ML service model training process, which significantly shortens the running time for verification. During the experiment, we fix $R=10$ restarts. The perturbation limitation has no significant impact on running time, as the running time remains consistent across different perturbation limits.


\noindent
\textbf{Impact on Unlearning Auditing Effect.} 
With larger perturbations, we augment the unlearning posterior difference, increasing the reconstruction similarity and enabling more information about the erased samples to be extracted. The trend is indicated in unlearning single samples (the green line) and multi-samples (the blue line) on all three datasets. Moreover, the results in the second row in \Cref{evaluation_of_noise} also clearly confirm our previous analysis: auditing for a single sample achieves a much better result than auditing for multiple samples, which is reflected in the significant gap between the two scenarios.

In TAPE, a larger perturbation of unlearned data makes data removal verification easier. In the third row in \Cref{evaluation_of_noise}, it is obvious that the verifiability increases as the perturbation limitation increases. When the perturbation limitation is less than $5$, it is hard to distinguish if the samples $D_{u,p}$ are unlearned because the remaining dataset contains a similar original sample $D_{u}$. However, when the perturbation limitation increases larger than $5$, the TAPE verifiability will greatly improve for genuine single-sample and multi-sample unlearning. MIB performs similarly when verifying the unlearning of backdoored multiple samples (B-MS). However, MIB fails to verify the unlearning of a single sample, as only one sample cannot backdoor the original ML service model.

\section{Summary and Future Work} \label{s_a_fw}

In this paper, we propose a TAPE scheme to investigate the auditing of unlearning effectiveness based on unlearning posterior differences, involving only the unlearning process. TAPE contributes a method to build unlearned shadow models to mimic the posterior difference quickly. Moreover, two strategies are introduced to augment the posterior difference, enabling the audit of unlearning multiple samples. The extensive experimental results validate the significant efficiency improvement compared with backdoor-based methods and the effectiveness of auditing genuine samples in both exact and approximate unlearning manners.

The auditing method proposed in this paper significantly addresses the limitations of existing unlearning verification methods. It effectively audits genuine samples for both exact and approximate unlearning methods in single-sample and multi-sample unlearning scenarios. Additionally, it eliminates the need for involvement in the original model training process. Future work could continue this line of inquiry, developing more efficient unlearning auditing methods to guarantee and support the right to be forgotten in MLaaS environments.

\begin{acks}
	This work is partially supported by Australia ARC LP220100453 and ARC DP240100955.
\end{acks}

\bibliographystyle{ACM-Reference-Format}
\balance
\bibliography{TAPE}

\appendix

\begin{table*}[h]
	\scriptsize
	\caption{An overview of machine unlearning auditing methods. 
	}
	\label{overview_of_auditing_method}
	\resizebox{\linewidth}{!}{
		\setlength\tabcolsep{3.pt}
		\begin{tabular}{c|cccccccc}
			\toprule[1pt]
			\multirow{2}{*} { \makecell[c]{\textbf{Unlearning} \\ \textbf{Auditing} \\ \textbf{Methods}} } & \multicolumn{2}{c} {\textbf{Involving Processes}  } & \multicolumn{2}{c} { \textbf{Auditing Data Type}} & \multicolumn{2}{c} {\textbf{Unlearning Methods}} & \multicolumn{2}{c} { \textbf{Unlearning Scenarios}}  \\
			\cmidrule(r){2-3}   \cmidrule(r){4-5} \cmidrule(r){6-7} \cmidrule(r){8-9}
			& \makecell[c]{{Original training} \\ {and unlearning	}  }  & \makecell[c]{{Only unlearning } \\ {process }  } & \makecell[c]{{Backdoored (marked)} \\ {samples	}  }    & \makecell[c]{{Genuine} \\ { samples	}  }   & \makecell[c]{{Exact} \\ {unlearning}  }    &\makecell[c]{{Approximate} \\ {unlearning}  } &   \makecell[c]{{Single} \\ {sample	}  }  &   \makecell[c]{{Multi} \\ {samples	}  }   \\ 
			\midrule
			MIB~\cite{hu2022membership} &\filledcircle & \emptycircle & \filledcircle&\emptycircle	&\filledcircle & \emptycircle &\emptycircle   	  & \filledcircle  \\
			Athena~\cite{sommer2022athena} &\filledcircle & \emptycircle & \filledcircle  &\emptycircle 	&\filledcircle &\emptycircle  &\emptycircle      & \filledcircle  \\
			Verify in the dark~\cite{guo2023verifying} &\filledcircle & \emptycircle & \filledcircle  &\emptycircle  &\filledcircle & \emptycircle &\emptycircle       & \filledcircle	  \\
			Verifi~\cite{gao2024verifi} &\filledcircle &\emptycircle & \filledcircle  &\emptycircle  	&\filledcircle & \emptycircle &\emptycircle       	  & \filledcircle  \\
			TAPE (Ours)	     &\emptycircle & \filledcircle & \emptycircle  & \filledcircle  	&\filledcircle  &\filledcircle & \filledcircle       	  & \filledcircle  \\
			\bottomrule[1pt]
	\end{tabular}}
	\begin{tabbing}
		\filledcircle: the auditing method is applicable; 
		\emptycircle: the auditing method is not applicable.
	\end{tabbing}
	\vspace{-2mm}
\end{table*}

\section{Additional Related Work Discussion} \label{different_with_existing}

\subsection{Machine Unlearning in the Web-Related Studies} 
Machine unlearning--the process of efficiently removing specific data influences from trained models--has been explored in diverse applications across Web-based systems, such as graph-based systems and personalized applications \citep{lin2024incentive,pan2023unlearning,zhu2023heterogeneous,wu2023gif}. In graph-based systems, \citeauthor{pan2023unlearning} \citep{pan2023unlearning} proposed an unlearning method to unlearn the graph classifiers with limited access to the original data, and \citeauthor{wu2023gif} \citep{wu2023gif} introduced a general strategy leveraging influence functions to efficiently remove specific graph data while preserving model integrity. In personalized applications, \citeauthor{lin2024incentive} \citep{lin2024incentive} introduced dynamic client selection with incentive mechanisms to enhance the federated unlearning efficiency, while \citep{zhu2023heterogeneous} extended federated unlearning to the heterogeneous knowledge graph, aiming to balance both privacy and model utility preservation. To achieve a better unlearning service, \cite{liu2024breaking} further explored the challenge of balancing privacy, utility, and efficiency and proposed a controllable unlearning framework to overcome this challenge.

\subsection{Difference from Existing Studies}  
Our TAPE approach is significantly different from existing unlearning verification methods \cite{hu2022membership,sommer2022athena,guo2023verifying,gao2024verifi} in terms of the involving processes, auditing data type, unlearning scenarios, and unlearning methods, as depicted in \Cref{overview_of_auditing_method}. First, the significant difference is that the auditing of our method only involves the unlearning process, while the backdoor-based methods must involve both the original training and unlearning processes to ensure the service model first learns the backdoor. Second, most existing auditing methods are based on backdooring techniques and need to backdoor or mark samples for verification \cite{hu2022membership,sommer2022athena,guo2023verifying,gao2024verifi}. As we analyzed in the above subsection, they can only validate the backdoored samples and are only applicable to the exact unlearning methods as exact unlearning methods guarantee the deletion from the dataset level. Our method does not mix any other data to the training dataset, and the auditing is based on the posterior difference, which is suitable for genuine samples in both exact and approximate unlearning methods. Third, backdoor-based auditing methods are only feasible for multi-sample unlearning scenarios because just using a single sample makes it hard to backdoor the model \cite{wang2019neural,lin2020composite,zeng2023narcissus,nguyen2020input}, hence failing to provide unlearning verification for a single sample.


\section{MLaaS Scenario and Threat Model} \label{threat_model}

Our problem is introduced in a simple machine unlearning as a service (MLaaS) scenario for ease of understanding. Under the MLaaS scenario, there are two main entities involved: an ML server that collects data from users, trains models, and provides the ML service, and users that contribute their data for ML model training. 

\noindent
\textbf{The ML Server's Ability.}
To uphold the ``right to be forgotten'' legislation and establish a privacy-protecting environment, the ML server is responsible for conducting machine unlearning operations. However, it is challenging to audit the unlearning effect for users to confirm that the unlearning is processed and prevent the spoof of unlearning from the ML server. In alignment with common unlearning verification settings \cite{hu2022membership,guo2023verifying}, we assume the ML server is honest for learning training but may spoof users for unlearning, i.e., it reliably hosts the learning process but may deceive users during unlearning operations by pretending unlearning has been executed when it has not. It is reasonable for the ML server to pretend to execute unlearning operations to avoid the degradation of model utility. Moreover, this assumption is more plausible than assuming the server will forge an unlearning update~\cite{thudi2022necessity}. Forging an unlearning update would require the server to simulate the disappearance of specified data and the corresponding resulting in model utility degradation, which demands significant effort without any benefit, making it an unlikely motivation. 

\noindent
\textbf{The Unlearning Users' Ability.}
We consider the scenario where the unlearning user has only black-box access to the ML service model, which is one of the most challenging scenarios \cite{salem2020updates,Hu2024sp}. In unlearning scenarios, the unlearning user possesses a local dataset, including the erased samples, which constitutes the entire training dataset for the ML service model \cite{warnecke2024machine,hu2024eraser}; however, the user has no access to the entire dataset. This just allows the user to query the model with their own data in a black-box access to obtain the corresponding posteriors and design the unlearning requests with specific data for unlearning verification purposes. 
Furthermore, we assume the unlearning user knows the unlearning algorithms, which is confirmed by both server and users, commonly used in other works \cite{hu2023duty}. However, even if the unlearning user knows the algorithms, without the remaining dataset, the user still cannot achieve the corresponding unlearning results of most unlearning algorithms. To relax the difficulty, we consider the unlearning user to be able to establish the same ML model as the current target ML service model with respect to model architecture. This can be achieved through model hyperparameter stealing attacks \cite{wang2018stealing,Seong2018towards,salem2020updates}. The unlearning user leverages this knowledge to simulate the unlearned shadow models and mimic the behavior of the ML service model based on the designed unlearning requests, thereby deriving the posterior differences necessary for training the reconstruction model to evaluate the unlearning effectiveness. 

 \section{Unlearning Data Perturbation (UDP) Algorithm} \label{UDP_algorithm}

 \Cref{Unlearned_d_p} demonstrates how to use the R restarts to find the satisfied perturbation for the unlearning data to augment the posterior difference for auditing.
 
 \begin{algorithm}[t]
 	\caption{Unlearning Data Perturbation (UDP)} \label{Unlearned_d_p}
 	\begin{small} 
 		\BlankLine
 		\KwIn{Trained model $\theta^*$, reconstruction model $\texttt{AE}$, unlearned data $X_u$, perturbation limit $\alpha$, local dataset $D_{local}$ }
 		\KwOut{The perturbed unlearning data, $X_u' = X_u + \Delta^{p}$} 
 		\SetNlSty{}{}{} 
 		\SetKwFunction{UDP}{\textbf{UDP}}
 		\SetKwProg{Fn}{procedure}{:}{end procedure}
 		\SetNlSty{}{}{} 
 		\Fn{\UDP{$\theta^*$, $\texttt{AE}$, $X_u$, $\alpha$, $D_{local}$ }}{
 			\For{$r \gets 1$ \KwTo $R$ restarts}{
 				$\Delta^{p}_{r} \gets \mathcal{N}(0,1)$  \hspace{4mm}    $\rhd$ Initialize random perturbation. \\
 				\For{$i \gets 1$ \KwTo $m$ optimization steps}{
 					$X_{u,i}^p \gets X_u + \Delta^{p}_r$  \hspace{0mm} $\rhd$ Add the perturbation to data. \\
 					$\theta_{\backslash (X_{u,i}^p)} \gets \theta^* - \frac{\epsilon}{n-1} \nabla \ell (X_{u,i}^p;\theta^*)$ $\rhd$  According to \Cref{shadow_model}.  \\
 					$\delta_{u,i}^p \gets \theta^*(D_{local}) - \theta_{\backslash (X_{u,i}^p)}(D_{local})$ $\rhd$ Calculate posterior difference according to \Cref{posterior_diff}.  \\
 					$\nabla \mathcal{L}_{\texttt{AE}} \gets \nabla \mathcal{L}_{\texttt{AE}} (\texttt{AE}(\delta_{u,i}^p) , X_{u,i}^p)$ $\rhd$  According to \Cref{perturb_loss}. \\			
 					$\Delta^{p}_{r} \gets \Delta^{p}_{r} - \eta \nabla \mathcal{L}_{\texttt{AE}} (\texttt{AE}(\delta_{u,i}^p) , X_{u,i}^p) $   \hspace{2mm} $\rhd$ Update perturbation with limitation $\|\Delta^{p}_{r}\|_{\infty} \leq \alpha $. \\
 				}
 			}
 			Choose the optimal $\Delta^p_r$ with minimal value in $\mathcal{L}_{\texttt{AE}}$ as $\Delta^{p*}$.\\
 			\Return $X_u' = X_u + \Delta^{p*}$
 		}
 	\end{small}
 \end{algorithm}

\section{The Verifier Training Process}  \label{verifi_train}

\begin{algorithm}[h]
	\caption{Verifier Model Training (VMT)} \label{verifier_trianing}
	\begin{small} 
		\BlankLine
		\KwIn{Reconstruction model $\texttt{AE}$, posterior differences $\delta$, local dataset $D_{local}$, unlearned dataset $D_u$ }
		\KwOut{The Verifier Model $\mathcal{V}$}  
		\SetNlSty{}{}{} 
		\SetKwFunction{VMT}{\textbf{VMT}}
		\SetKwProg{Fn}{procedure}{:}{end procedure}
		\SetNlSty{}{}{} 
		\Fn{\VMT{$\texttt{AE}$, $\delta$, $D_{local}$, $D_u$}}{
			Initialize a verification dataset $D_{veri.}$ \\
			\For{$x_u$ in $D_u$, $x_i$ in $D_{local} \backslash D_u$ }{
				$D_{veri.}$ adds the positive sample ($\texttt{AE}(\delta_{\backslash x_u}), x_u; 1$) \\
				$D_{veri.}$ adds the negative sample ($\texttt{AE}(\delta_{\backslash x_u}), x_i; 0$) \\
			}
			Initialize a Verifier model $\mathcal{V}$ \\
			Train $\mathcal{V}$ on the constructed $D_{veri.}$ using a cross entropy loss  \\
			\Return the trained $\mathcal{V}$
		}
	\end{small}
\end{algorithm}


This Verifier aims to identify if the recovered samples are unlearned samples. Specifically, we first construct a verification dataset $D_{veri.}$. For each instance in the unlearned dataset, and the reconstructor model reconstructs based on the posterior difference of the instance, and we set the corresponding label equal to 1. We add it as the postive sample ($\texttt{AE}(\delta_{\backslash x_u}), x_u; 1$) into $D_{veri.}$ For each instance in the local dataset that is not part of the unlearned dataset, we set a negative label for the instance and the reconstructed sample pair, i.e. ($\texttt{AE}(\delta_{\backslash x_u}), x_i; 0$). These samples are added to the verification dataset too. A verifier model is then initialized and trained on this constructed dataset using a cross-entropy loss. The Verifier model training algorithm is presented in \Cref{verifier_trianing}. 

\section{Metrics and Requirements for Auditing} \label{detailed_metrics}

\begin{table*}[t]
	\scriptsize
	\caption{Overall Evaluation Results on MNIST, CIFAR10, STL-10, and CelebA. 	\vspace{-2mm}}
	\label{tab_total}
	\resizebox{\linewidth}{!}{
		\setlength\tabcolsep{4.pt}
		\begin{tabular}{c|cccccccccccc}
			\toprule[1pt]
			\multirow{2}{*} { \makecell[c]{\textbf{Single-Sample} \\ \textbf{Unlearning Auditing}} } & \multicolumn{3}{c} {MNIST, $\text{\it ESS}=1$}& \multicolumn{3}{c} {CIFAR10, $\text{\it ESS}=1$} & \multicolumn{3}{c} {STL-10, $\text{\it ESS}=1$}  & \multicolumn{3}{c} {CelebA, $\text{\it ESS}=1$} \\
			\cmidrule(r){2-4}   \cmidrule(r){5-7} \cmidrule(r){8-10} \cmidrule(r){11-13}
			& Original & MIB \cite{hu2022membership}  	 & TAPE	  & Original	& MIB    & TAPE	 & Original	 &   MIB    & TAPE  & Original	 &   MIB    & TAPE \\
			\midrule 
			Running time (s)  	  & 620 &  	637								& \textbf{143}  		         &651  &  672	& \textbf{135}	 	&781 		& 815 & \textbf{79.81}  &1546 		& 1622 & \textbf{32.76} \\
			Model Utility (Acc.)	 &\textbf{99.14\%}       & 98.31\%   &\textbf{99.14\%}         & \textbf{81.62\%} 	  & 79.45\%   & \textbf{81.62\%}   &\textbf{68.99\%} & 67.54\% & \textbf{68.99\%}  & \textbf{96.93\% } & 96.05\%  & \textbf{96.93\% }  \\
			Rec. Sim.  		 	& -  		 & -  					&\textbf{0.965}   						 			& - & - & \textbf{0.974} & -  	   &-& \textbf{0.175}  & -  	   &-& \textbf{0.977} \\
			Unl. Verifiability     &  $0.00\%$   & $0.00\%$  &\textbf{99.43\%}    &   $0.00\%$ &  $0.00\%$ & \textbf{98.76\%}   &  $0.00\%$  &  $0.00\%$   & \textbf{84.00\%}  &  $0.00\%$ &  $0.00\%$ & \textbf{97.64\%} \\
			\midrule[0.12em]
			\multirow{2}{*} {\makecell[c]{\textbf{Multi-Sample} \\ \textbf{Unlearning Auditing}}} & \multicolumn{3}{c} {MNIST, $\text{\it ESS}=20$}& \multicolumn{3}{c} {CIFAR10, $\text{\it ESS}=20$} & \multicolumn{3}{c} {STL-10, $\text{\it ESS}=2$} & \multicolumn{3}{c} {CelebA, $\text{\it ESS}=20$} \\
			\cmidrule(r){2-4}   \cmidrule(r){5-7} \cmidrule(r){8-10} \cmidrule(r){11-13}
			& Original & MIB \cite{hu2022membership}   	 & TAPE & Original & MIB   & TAPE	& Original	 &   MIB   & TAPE & Original	 &   MIB   & TAPE  \\
			\midrule 
			Running time (s)   & 613  & 638    &  \textbf{113}    & 644   & 673 			& \textbf{113} 	 & 781	& 809  & \textbf{74.90}   & 1570	& 1663  & \textbf{21.43}  \\
			Model Acc.	     & \textbf{99.05\%}   & 98.73\%    & \textbf{99.05\%}    & \textbf{81.62\%}    & 79.13\%    & \textbf{81.62\%} & \textbf{68.99\%} &  67.26\% &  \textbf{68.99\%}  &  \textbf{97.01\%} &  96.88\%  &  \textbf{97.01\%} \\
			Rec. Sim.  			    &-   			& -  			& \textbf{0.933}  			&-& -									& \textbf{0.973}					 	&- & - & \textbf{0.174}  &- & - & \textbf{0.970} \\
			Unl. Verifiability   	  & $0.00\%$   & $0.00\%$ & \textbf{98.67\%}     & $0.00\%$ & $0.00\%$ & \textbf{97.44\%}  	& $0.00\%$ & $0.00\%$ & \textbf{84.40\%} &  $0.00\%$ &  $0.00\%$  &  \textbf{94.57\%} \\
			\bottomrule[1pt]
	\end{tabular}}
\end{table*}


\noindent
\textbf{Data Removal Verifiability.} 
Existing backdoor-based unlearning verification methods can only provide the data removal verifiability based on the backdoor attack success rate \cite{guo2023verifying,hu2022membership}. We also train a Verifier (a classifying model) to identify the reconstructed data of the unlearned samples and the reconstructed data of the samples that still remain. We propose Verifiability to evaluate the accuracy of the Verifier, which calculates the correct classifying rate as 
\begin{equation} \label{verifiability_def}
	\textbf{Verifiability: } \hspace{4mm}  V =  \frac{ 1}{m}\sum_{x_u \in D_u} \mathbb{I}( \text{Verifier}(\delta_u, x_{u})=1),  
\end{equation}
where $m$ is the size of the unlearned dataset $D_{u}$ and $\mathbb{I}$ is the indicator function that equals 1 when its argument is true ($\text{Verifier}(\delta_u, x_{u})=1$) and 0 otherwise.

\section{Additional Experiments} \label{additional_exp}

\subsection{Overview Evaluation of TAPE} \label{overall_eval_app}

We demonstrate the overview evaluation results of different unlearning auditing methods on MNIST, CIFAR10, STL-10 and CelebA, presented in \Cref{tab_total}. The upper half of \Cref{tab_total} demonstrates the evaluations of the single-sample unlearning auditing, and the lower half of \Cref{tab_total} presents the evaluations of the multi-sample unlearning auditing. The bolded values indicate the best performance among the compared methods. We fill a dash when the method does not contain the evaluation metrics.

\noindent
\textbf{Setup.}
We measure auditing methods based on the four above-introduced evaluation metrics in single-sample and multi-sample unlearning scenarios. In single-sample verification, the Erased Sample Size ({\it ESS}) is equal to 1 and $\text{\it ESS}=20$ for the multi-sample scenario. On STL-10, we set $\text{\it ESS}=2$ for the multi-sample scenario, as STL-10 only contains 5000 training samples, which is much smaller than other datasets. The evaluation here is tested based on the retraining-based unlearning method SISA \cite{bourtoule2021machine}. To better illustrate the functionality preservation and efficiency, we record the performance of solely training the original model, shown as ``Original'' in \Cref{tab_total}.

\noindent
\textbf{Evaluation of Efficiency.}
Since TAPE does not involve the original model training process, it consumes much less running time than MIB and ``Original''. The ``Original'' is training the original model before unlearning, and the MIB method needs to backdoor the model during the initial model training process before unlearning. Specifically, TAPE achieves more than $4.5\times$ speedup in efficiency on MNIST, $5\times$ speedup on CIFAR10, $10\times$ speedup on STL-10, and $50\times$ speedup on CelebA. On CelebA, the best speedup is up to $75\times$.

\noindent
\textbf{Evaluation of Functionality Preservation.}
The effect of functionality preservation is measured by model accuracy. In both single-sample and multi-sample unlearning auditing, our TAPE always achieves better functionality preservation than MIB. The highest accuracy preservation is around $2\%$, achieved on CIFAR10. The reason is that the MIB method needs to mix backdoored samples into the training dataset, and the backdoored samples with modified labels will negatively influence model utility. On the contrary, the TAPE scheme is independent of the original model training process; hence, our method will not influence the model utility of the original ML service model, keeping the same model accuracy as ``Original'', demonstrating better functionality preservation.

\noindent
\textbf{Evaluation of Unlearning Auditing Effect.}
We use reconstruction similarity to measure how much information about the specified samples is unlearned. The MIB method is unable to provide such an assessment of unlearned information for evaluation of unlearning effectiveness. Hence, we fill a dash of MIB in this metric. Reconstruction for a single sample always achieves better results than for multiple samples, which confirms our previous analysis and existing works \cite{salem2020updates,balle2022reconstructing}. The unlearned posterior difference of a single sample contains more information about such a sample than a posterior difference of multiple samples, as information from multiple samples is interwoven together in one posterior difference.

To align with existing unlearning verification methods, we propose the verifiability metric to evaluate the data removal status, which is defined in \Cref{verifiability_def}. Since the erased sample size is small and only genuine unlearned samples are evaluated in this experiment, the MIB cannot successfully verify the unlearning of any genuine samples in \Cref{tab_total}. In both single-sample and multi-sample unlearning scenarios, TAPE provides effective data removal verification (accuracy larger than $95\%$ on MNIST, CIFAR10, and CelebA). Moreover, data removal status verification for a single sample always achieves better results than for multiple samples.

\subsection{Impact on Efficiency of Erased Samples Size ($\text{\it ESS}$)}  \label{imp_of_efficiency_ESS}

\noindent
\textbf{Impact on Efficiency.} 
The main components of the running time of TAPE are building unlearned shadow models and training the reconstructor. The running time of these two processes is highly related to the size of the user's local dataset. In our experiments, we randomly select $0.5\%$ samples on MNIST and CIFAR10 and choose $0.06\%$ samples on CelebA as the local dataset.

\begin{figure}[t]
	\centering
	\subfloat{    
		\includegraphics[scale=0.4]{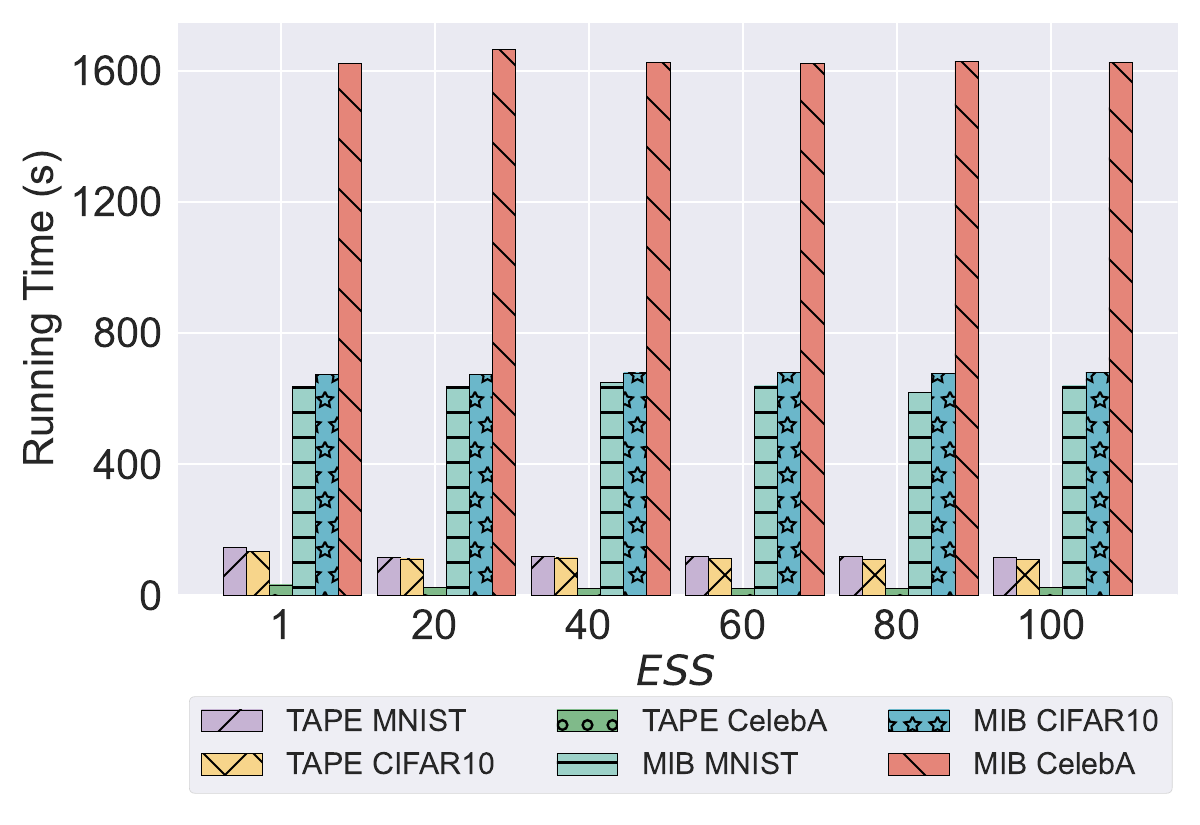}
	}
	\vspace{-1mm}
	\caption{Running time about different $ESS$.}
	\vspace{-2mm}
	\label{evaluation_of_running_time} 
\end{figure}

\Cref{evaluation_of_running_time} shows the running time of TAPE and MIB on the three datasets. The running time of TAPE has no significant relationship with the $\text{\it ESS}$ because the running time of TAPE (shadow model building and reconstructor training) highly depends on the size of the user's local dataset. For MIB, the running time has no obvious variations when $\text{\it ESS}$ increases. This is because the MIB verification preparation is accompanied by the original model training, which is heavily related to the size of training datasets. 
TAPE has a much more efficient running time compared with MIB, as TAPE is independent of the original model training.

\end{document}